\documentclass[12pt]{article}
\usepackage{booktabs}
\usepackage{pdflscape}
\usepackage{afterpage}
\usepackage{capt-of}
\usepackage[pdftex]{graphicx}
\usepackage{amssymb}
\usepackage{subfigure}
\usepackage{caption}
\usepackage{float}
\usepackage{natbib}
\usepackage{comment}
\usepackage{xargs}
\usepackage{anysize}
\usepackage{lipsum}
\usepackage[utf8x]{inputenc}

\marginsize{2.0 cm}{2.0 cm}{2.0 cm}{2.0 cm}

\begin{document}

%\begin{frontmatter}
%
%%% Title, authors and addresses
%
%\title{How to be a Made Man: a Social Network Analysis of the Internal Structure of the Sicilian Mafia}
%
%\author[label1]{Michele Battisti}
%\author[label1]{Andrea Mario Lavezzi}
%\author[label2]{Roberto Musotto}
%\address[label1]{Department of Law, University of Palermo}
%\address[label2]{CCJS, School of Law, University of Leeds}
%
%\date{today}
\title{Organizing Crime: an Empirical Analysis of the Sicilian Mafia\thanks{We thank Francesco Calderoni, Valentino Dardanoni, Vincenzo Militello, Paolo Pin, and seminar participants at Brunel University and University of Palermo for comments and suggestions. We thank Marco Manno for helping us in obtaining the data. Usual disclaimer applies.}}

\author{Michele Battisti\thanks{Department of Law, University of Palermo, Piazza Bologni 8 - 90134 Palermo, Italy. Email: michele.battisti@unipa.it.} \and Andrea Mario Lavezzi\thanks{Department of Law, University of Palermo, Piazza Bologni 8 - 90134 Palermo, Italy. Email: mario.lavezzi@unipa.it.} \and Roberto Musotto\thanks{Department of the Premier and Cabinet, Western Australia. Email: Roberto.Musotto@dpc.wa.gov.au.}}
\date{\today\\[1cm]}
\maketitle
\vspace{-2cm}
\begin{abstract}
In this article we study the organizational structure of a large group of members of the Sicilian Mafia by means of social network analysis and an econometric analysis of link formation. Our mains results are the following. i) The Mafia network is a small-world network adjusted by its criminal nature, and is strongly disassortative. ii) Mafia bosses are not always central in the network. In particular, consistent with a prediction of \citet{baccara2008organize}, we identify a ``cell-dominated hierarchy" in the network: a key member is not central, but is connected to a relative with a central position. iii) The probability of link formation between two agents is higher if the two agents belong to the same \textit{Mandamento}, if they share a high number of similar tasks, while being a ``boss" reduces the probability of link formation between them. iv) The probability of link formation for an individual agent is higher if he is in charge of keeping connections outside his \textit{Mandamento}, of collecting protection money and or having a directive role, while age has modest role. These results are interpreted in the light of the efficiency/security trade-off faced by the Mafia and of its known hierarchical structure.

\end{abstract}

%\begin{keyword}

\section{Introduction\label{secIntro}}
Mafias are criminal organizations operating in various countries, having in some cases long histories. Among the most well-known cases, one can mention \textit{Cosa Nostra}, the Sicilian Mafia, and the Russian Mafia (see, respectively, \citealp{gambetta1996sicilian} and \citealp{varese2001russian}, for thorough analyses). What distinguishes a Mafia from a generic organized group of criminals is that a Mafia is not only involved in illegal activities (e.g. smuggling of drugs and arms, counterfeiting, extortion, etc.) but it also provides a specific service typical of a State,  namely \textit{protection} \citep{federico2010organized}.

This paper proposes an analysis of the internal organization of the Sicilian Mafia viewed as a \textit{network}, i.e. a set of agents connected by a set of links. It does so by applying two tools of empirical analysis, social network analysis and dyadic regression analysis, to an original dataset of a large group of \textit{Cosa Nostra} members, built from judicial sources on the investigation named \textit{Operazione Perseo}, a recent anti-Mafia operation that brought to the arrest of a large number of \textit{mafiosi}.

As pointed out by a large literature on organized crime, a criminal organization tries to perform its various activities struggling between the need to be efficient and the necessity to hide its operations to the legal authorities. This trade-off is traditionally denoted as the efficiency/secrecy trade-off (see, e.g., \citealp{morselli2007efficiency}). Specifically, the illicit trades carried out on daily bases and the fundamental goal of providing  protection, are likely to require an efficient internal organization, based for example on fast circulation of information, on some form of division of labor, and on the interaction among the \textit{Cosa Nostra} syndicates that rule over different territorial areas. In fact, a distinguishing feature of the Sicilian Mafia is a rigid definition of portions of territory, named \textit{Mandamenti}, over which one or more families have power. To the best of our knowledge, this article is the first one that tries to shed light on the organization of a Mafia by resorting to the joint use of the social network analysis and of the estimation of dyadic regressions.

%Understanding the internal organization of a Mafia could be considered a goal not only interesting in itself, but also potentially useful to inform legal authorities on how to devise optimal strategies to attack, weaken, and possibly disrupt these organizations.

%recently advocated by \citet{graham2019} as a useful econometric method for analyzing unobserved agent-specific heterogeneity to identity relevant correlations among individual \textit{mafiosi}'s characteristics and the network structure. 

Our results are the following. The social network analysis shows that: i) the network of agents involved in \textit{Operazione Perseo} network is a small-world network. That is, it has a short diameter and a high clustering coefficient. However, with respect to (simulated) comparable licit small-world networks, the clustering coefficient and the diameter are, respectively, significantly smaller and higher. We argue that this can be interpreted as a way to increase the security of the organization. ii) The network is strongly disassortative, i.e. individuals with a high number of links tend to be connected with little connected individuals, in contrast to other small-world networks which are typically assortative. We argue that this is an implication of the hierarchical structure of \textit{Cosa Nostra}. iii) Consistent with a prediction of the theoretical model of \citet{baccara2008organize}, we find that a ``cell-dominated hierarchy" characterizes the network. In fact, a key member of the network is not central in the network, as intuition can suggest. On the contrary, a much higher central position in the network, identified by a high number of links, is held by a relative of that key agent. The econometric analysis shows that: iv) The probability of link formation between two agents is higher if the two agents belong to the same \textit{Mandamento}, and the higher is the number of similar tasks performed by the agents, and is lower if the two agents are ``bosses". v) The probability of link formation for an individual agent is higher if he is in charge of keeping connections outside his \textit{Mandamento}, or of collecting protection money, or if he is a boss, while age has a moderate positive effect.

The paper is organized as follows: Section \ref{secLiterature} discusses the related literature; Section \ref{secIntStruct} introduces the known organizational features of \textit{Cosa Nostra}; Section \ref{secData} describes the dataset; Section \ref{secResults} presents the main results of our analysis and discusses the implications for the understanding of the internal organization of a mafia network; Section \ref{secConcl} concludes.

\section{Related Literature\label{secLiterature}}

Organized crime, in particular the Sicilian Mafia, has traditionally attracted the interest of sociologists (e.g. \citealp{gambetta1996sicilian}), criminologists (e.g. \citealp{paoli2004italian}) and, more recently, of economists (see, e.g., \citealp{acemoglu2020}, on Mafia origins, \citealp{pinotti2015economic}, on Mafias and economic growth, \citealp{balletta2019extortion}, on extortion).

The structure of organized crime groups was analyzed through social network analysis in other works. For example, \citet{varese2012mafias,varese2012structure} study a Russian criminal group operating in Italy highlighting the topological characteristics (e.g. degree distribution, centrality) of its network structure, while \citet{calderoni2012structure} offers a similar analysis of a drug trafficking network belonging to \textit{'Ndrangheta}, the criminal organization from the Italian region of Calabria. \citet{calderoni2012structure}, in particular, finds that the criminals most involved in the drug trafficking operations are not always the most central in the network. Finally, \citet{catino2022network} also focuses on the \textit{'Ndrangheta} and studies the network created by interfamily marriages, a tool strategically utilized by the criminal organization to foster trust across families. Although these studies share with the present work the estimation of the topological properties of the criminal network, they are not based on the Sicilian Mafia and do not utilize dyadic regression analysis as the present work.

The network structure of the Sicilian Mafia has been analyzed in some recent works. \citet{agreste2016network} study a large  \textit{Cosa Nostra} network operating in North-Eastern Sicily, on the basis of several court documents. They identify two types of networks: one based on wiretapped conversations, and one based on different sources of information (bank transactions, co-offences, etc.), and find that the topological properties of the two networks are remarkably different. \citet{cavallaro2020disrupting} study the operation \textit{Montagna}, which disrupted a \textit{Cosa Nostra} group operating in the province of Messina around 2007. This work points out that targeting agents with the largest betweenness centrality is an effective policy to disrupt criminal organizations. \citet{calderoni2020robust} consider the same dataset of \citet{cavallaro2020disrupting}, and study the problem of link prediction, showing that measures such as the Katz centrality score performs well in predicting links. Finally, \citet{tumminello2021anagraphical} study a large dataset from recent investigations on the Sicilian Mafia. They find that members of the same Mafia syndicate are likely to live in the same territory, that Mafia members become more and more specialized as their criminal career proceeds and, finally, that women are often crucial in connecting different syndicates. Although some of these works proposes a network analysis similar to this work, with the exception of \citet{tumminello2021anagraphical} who study bipartite networks, none of these works estimate dyadic regressions.

Finally, two recent works in economics are relevant for our work. On the empirical side, \citet{mastrobuoni2012organized} study the correlation of a vast array of individual characteristics of members of the American Mafia with their position in its network structure. We also provide an econometric analysis that correlates individual characteristics to the network structure, but we estimate dyadic regressions and not OLS regressions as in \citet{mastrobuoni2012organized}. On the theoretical side, \citet{baccara2008organize}, although not explicitly addressed to the Sicilian Mafia, formalize the problem of a criminal organization facing an efficiency/security trade-off. In particular, \citet{baccara2008organize} identify the optimal information structures that a criminal organization can choose to solve the trade-off. Interestingly, \citet{baccara2008organize} show that, under a vast array of possibilities, the optimal structure of a criminal organization is characterized by a hierarchy, in which some agents act as \textit{information hubs} by possessing information on a number of other agents who do not possess information on each other. With respect to this work, we will show that some features of the empirical network criminal structure from the \textit{Perseo} operation are consistent with the theoretical predictions of \citet{baccara2008organize}.

\section{On the Internal Structure of \textit{Cosa Nostra}\label{secIntStruct}}

The criminal organization known as \textit{Cosa Nostra} originated in the Italian region of Sicily in the nineteenth century.\footnote{See, e.g., \cite{gambetta1996sicilian} and and \citet{lupo2004storia} for detailed accounts.} At the time of the Italian unification, it was defined in a parliamentary enquiry as ``an industry of violence" \citep{franchetti1877sicilia} and, over time, it become the epitome of Mafia-type criminal organizations, i.e. organizations aiming at providing extralegal governance over a specific territory. In particular, the crucial element characterizing the Sicilian Mafia is the production and distribution in the underworld and in the legal sphere of a specific commodity: protection (from other criminals, from competitors, etc.).\footnote{See in particular \citet{gambetta1996sicilian} and \citet{varese2011mafias}.} To do so, it must achieve monopolistic control of the provision of protection over a specific area, as extralegal governance cannot be provided by more than one authority on a territory \citep{schelling1971business}.

Recruitment into the organization typically takes place by a ritual, in which the new member takes a solemn oath of loyalty to the organization (see, e.g., \citealp[p. 146]{gambetta1996sicilian}, and \citealp[p. 67]{catino2019mafia}). The new member is introduced by a former member, and the oath is taken in front of other \textit{men of honour}, as members of \textit{Cosa Nostra} are denoted. A family tie is very relevant in this context as: ``to be born into a mafioso family is considered as sufficient guarantee of eligibility" \citep[p. 147]{gambetta1996sicilian}.

 \textit{Cosa Nostra} has a hierarchical structure as its internal organization has multiple levels (see, e.g., \citealp[pp. 153-159]{catino2019mafia}). Starting from the lowest, the first level is represented by a \textit{soldato} (``soldier"). Ten or more soldiers are headed by a \textit{capo-decina} (``head of ten"). All the soldiers are also part of a family, denoted as \textit{cosca}. The family is controlled and directed by a family leader. In some territories, like in the city of Palermo, three or more families rule over a \textit{Mandamento}, i.e. a portion of the city territory.\footnote{According to recent investigations, the city of Palermo is currently divided into eight \textit{Mandamenti}. See Appendix \ref{secAMand}.} A \textit{Mandamento} is controlled by a \textit{Capo-mandamento}, who is a member of the \textit{Commissione Provinciale} (``Provincial Commission"), together with other \textit{Capi-mandamenti}. On top of all provincial commissions, a \textit{Cupola} rules, in which all the provincial representatives take part. The \textit{Cupola} has the directive role of the entire organization \citep[p. 157]{catino2019mafia}. In particular, the \textit{Cupola} takes the most important decisions such as assassinations of high-profile individuals like magistrates or politicians. Currently, the highest levels of the Sicilian Mafia have been dismantled by police operations, although evidence exists, also from the \textit{Perseo} operation, that attempts have been recently made to rebuild the higher-level bodies such as the Provincial Commission of Palermo \citep[p. 158]{catino2019mafia}.

\section{The Dataset\label{secData}}

Data come from judicial evidence of the so-called \textit{Operazione Perseo}, a police operation that took place in December 2008 to disrupt a large \textit{Cosa Nostra} group operating mostly in Palermo, that tried to establish a new \textit{Cupola}. The operation covers a two-year period, from late 2006 to 2008. \textit{Operazione Perseo} brought to the arrest of 99 people, the largest number of arrests with respect to the most recent police operations against \textit{Cosa Nostra} in the area of Palermo (see Table \ref{tablePoliceOp}). In particular, \textit{Operazione Perseo} is the second biggest trial on Mafia after the \textit{Maxiprocesso} (``maxi-trial") of 1986 for number of arrests.\footnote{The \textit{Maxiprocesso} was the first, and so far unparalleled, large-scale trial against the Sicilian Mafia, and brought to the arrest of 474 \textit{mafiosi}.} \textit{Operazione Perseo} hit almost all the families and turfs in the Palermo province offering an ideal picture to investigate how they were networked.

 \begin{table}[ptb]
  \centering
 \begin{tabular}{l|l|c}
 Year & Operation Name & Arrests \\ \hline
 2016 & Monte Reale & 16 \\ 
 2016 & Brasca-quattro.zero & 62 \\
 2016 & Cicero & 9 \\ 
 2015 & Jafar & 7 \\ 
 2015 & Stirpe & 5\\ 
 2015 & Verbero & 39 \\ 
 2015 & Panta Rei & 39 \\ 
 2014 & Apocalisse 1 and 2 & 97\\ 
 2014 & Grande Passo 1, 2, 3 and 4 & 28 \\ 
 2014 & Eden 1 and 2 & 16 \\ 
 2013 & Nuovo Mandamento & 37\\ 
 2013 & Zefiro & 18 \\ 
 2012 & Atropos & 41 \\ 
 2011 & Pedro & 28 \\ 
 2011 & Hydra & 16 \\ 
 2008 & \textbf{Perseo} & \textbf{99} \\ 
 2008 & Old Bridge & 90 \\ 
 2006 & Gotha & 52 \\ 
 2005 & Grande Mandamento & 46 \\
% 1994 & Petrov & 74 \\ 
% 1986 & Maxiprocesso & 474 \\ \hline
% 1985 & Pizza connection & 19 \\ \hline
% 1968 & Strage di Ciaculli & 117 \\ \hline
% 1925-1929 & Mori arrests & 11000 \\ \hline
% 1898-1900 & Sangiorgi Reports & 218 \\ \hline
% 1883 & La Fratellanza & 200 \\ \hline
% 1862 & & 13 \\

% Total & around 31 investigations &  12860 arrests\\ \hlin
 \end{tabular}
 \caption{Police operations against the Mafia in the province of Palermo, 2005-2016}
  \label{tablePoliceOp}
 \end{table}
 
First of all, we provide a detailed description of the main court document that has been analyzed, explaining why it allows for a less biased representation of the network of interest than other sources. Subsequently, we explain how the information was coded into a social network representation.

\subsection{The Data Source \label{secDataSources}}
The document chosen to construct the dataset is the \textit{fermo di indiziati di delitto}, disciplined by Art. 384 of the Italian penal code. It represents the arrest warrant, issued by the magistrates to support the arrest of individuals under investigation. Among all the acts, papers, media and documents from a trial, the arrest warrant is the preferred one for the amount and the quality of the information it contains \citep{musotto2021}. It is the first act in the Italian criminal procedure law that discloses the whole investigation and, most of the time, the last document of every investigation, implying the notification of its end. 
 
%@article{musotto20204,
 % title={From evidence to proof: social network analysis in Italian Criminal Courts of Justice},
  %author={Musotto, Roberto},
  %journal={SPECIAL COLLECTION ON ARTIFICIAL INTELLIGENCE},
  %pages={38},
  %year={2020}
%} 
 
The most important feature of the warrant is that the Public Prosecutor is legally obliged to present every proof against and in favor of the person that is currently under investigation. This means that technically the Prosecutor does not have any real power to select proofs.\footnote{The only evidence the Prosecutor is actually allowed to discard are the one that are blatantly redundant (e.g. two identical wiretaps of the same meeting) or not relevant for the specific investigation (i.e. evidence from the suspect while not engaged in suspicious or criminal activity).} Therefore, it is the most accurate document compiled by the authorities regarding certain criminal activities. This is not the case for other documents collected during the trial where judges, the jury, public prosecutors and lawyers decide if a certain piece of evidence should be admitted for judgement and then evaluated. 
 
 For the purposes of this paper, therefore, it is the best document that can be employed to reconstruct the network of connections (links) among the members of the criminal group (nodes), compared to other possible reconstructions of the network obtained from other court documents \citep{berlusconi2016link}. For example, as pointed out by \citet{berlusconi2013all}, the sentence is less apt for such an analysis than the arrest warrant in which proofs have not yet been filtered through the procedure of the judgement. In addition, from the point of view of network analysis, all the observable information about connections among individuals is gathered, with the help of wiretap records, audio and video collections, etc. The information which is collected should be, therefore, the least biased in terms of missing data and self-selection.\footnote{It goes without saying that a pure representative sample of the network could hardly be obtained as the relevant universe, i.e. the complete set of nodes and of their connections, is simply not observable given the illegal nature of the organization and of its activities. This is especially true for a criminal organization such as the Sicilian Mafia, which is strictly devoted to secrecy and to the obedience to the code of \textit{omerta'} \citep[p. 121]{gambetta1996sicilian}. Indeed, other works  (e.g. \citealp{scaglione2011reti,campana2012listening}) utilized the arrest warrant as a starting point for social network analysis.} 

However, some pitfalls related to this type of data collection still remain. First of all, although the information from the arrest warrant is abundant in many respects, the precise boundaries of the network might not be completely identified. In the specific case of \textit{Operazione Perseo} the problem is that, even if there might not be omissions in the data, the investigation focused only on one part of the Mafia in or around the area of Palermo. For example, it is not possible to claim that all the members in that area were captured. Also, all the relational ties that lead outside this geographical area are partially excluded from the analysis. This is the case, in particular, of the relationships between the Mafia network in Palermo and the one in Trapani.\footnote{Specifically, in the arrest warrant there are constant references to the major boss of \textit{Cosa Nostra}, Matteo Messina Denaro, who is from the province of Trapani. The links with that part of the organization, however, are treated in the investigation as if they are outside the network. The relationships are recorded, but they are not deeply analyzed.}

In addition, from the perspective of the social network analysis it is in almost all cases not possible to clearly identify the \textit{direction} of the link, i.e. given a link between agents $i$ and $j$ it is not possible to identify which agent started the relationship. Also, from the document examined it is not possible to specify the ``weight" of a link, i.e. the intensity of the relationship between two nodes, e.g. the frequency with which they used to meet or interact.\footnote{See, e.g., \citet[p. 21]{jackson2008social} on directed and weighted links.}

\subsection{The Network \label{secRecNet}}
 The dataset on the network was built through a complete analysis of the judicial text. All personal and sensitive information are not reported in what follows in order to protect the privacy of anyone involved in the investigation.

The network is a collection of nodes and links extracted from the arrest warrant. The individuals involved in the investigation are the primary set of nodes that we consider to reconstruct the network, and correspond to the ninety-nine people arrested. Other seventy-three people appear in the investigation as linked to the ninety-nine arrested.
%\footnote{Most of the other individuals involved have been arrested during other anti-Mafia operations such as \textit{Grande Mandamento} or \textit{Addio Pizzo}.}
 Overall, the network we reconstructed contains 172 agents, 99 of which were directly affected by the operation and arrested. However, since the information in the arrest warrant is detailed only for the agents that were arrested, the analysis will be based on the group of 99 agents, and will cursorily refer to the larger network of 172 agents. For simplicity, in what follows the network of 99 agents will be referred to as the \textit{Perseo} network.

For each agent, information has been collected on the following characteristics: name and gender,\footnote{Although women are not formally allowed into \textit{Cosa Nostra}, they can give support to the family in its daily activities \citep[p. 36]{tumminello2021anagraphical}. In the set of 99 agents, however, no women were included.} place and date of birth, place of residence (city and address), occupation (if available), \textit{Mandamento} affiliation,\footnote{See Figures \ref{fig:diaMandamPa} and \ref{fig:diaMandamPaPr} for the list and localization of \textit{Mandamenti} in the city and province of Palermo, according to the official investigations of D.I.A. (\textit{Direzione Investigativa Antimafia}), a body operating at the Italian Ministry of Internal Affairs, specialized in investigations on organized crime. See \citet{dia2016}.} the indication on whether, according to the investigators, the agent had a directive role in the organization, the agent's tasks as identified by the prosecutors.\footnote{The initial section of the arrest warrant lists the crimes ascribed to the arrested persons (\textit{capi d'imputazione}). The Prosecutor in this section also provides a brief description of each arrested persons, indicating whether he had a directive role, his principal tasks in the organizations and a list of the other individuals with whom, according to the investigation, he had relevant interactions. The arrest warrant indicates at least one task for all members of the \textit{Perseo} network.  \label{footIntroDoc}} In particular, eight different tasks were identified, and were coded as: \textit{LinkingIn}, if the agent was responsible of keeping connections inside his own \textit{Mandamento}; \textit{LinkingOut}, if the agent was in charge of keeping connections outside his own \textit{Mandamento}; \textit{Meeting}, if the agent was supposed of attending meetings with other members of the organization in person;\footnote{\citet{calderoni2019nature} specifically focus on the social network structure based on attending meetings, for a group of members of the \textit{'Ndrangheta}, the criminal organization from the Italian region of Calabria.} \textit{PubProc}, if the agent was involved in influencing the adjudication of public works through public procurement; \textit{Pizzo}, if the agent was in charge of collecting protection money (denoted as \textit{Pizzo} in Sicilian); \textit{Guns} if the agent had the task of smuggling arms; \textit{Drugs} if the agent had the tasks of of drug production and smuggling; \textit{host} if the agent provided accommodation to other members.\footnote{Most of the agents were associated to more than one task. The task \textit{host} was indicated for only one member of the set of 99 agents, so we will exclude it from our main empirical analysis.} Finally, we considered the latitude and longitude of each member's address, in order to geo-localize the agent in the territory. This will allow to compute the geographical distance among the agents, which will bu utilized in the econometric analysis.\footnote{In a companion paper, \citet{battistiEtAlSpatial2022}, we provide a detailed spatial network analysis of the \textit{Perseo} network, where the geographical distance among the agents is the key variable.}

To identify the existence of a link between any two nodes of the network, we collected every piece of information in the court document on a connection between any two agents. This brought to the collection of 1410 pieces of evidence on the connections among the nodes of the network. 

Subsequently, we utilized these pieces of information to define the links among the agents. In particular, we assumed that a link between two agents exists if at least one of these conditions is fulfilled: i) the prosecutors pointed it out in the first section of the arrest warrant (see Footnote \ref{footIntroDoc});\footnote{To be more precise, the first section of the arrest warrant contains pieces of information like the following for agent 110 (a random number to preserve anonimity, see Footnote \ref{footRandNum}) : `` ... he substituted [agent 77] in the direction of the Mafia family  ... and 	kept contacts with members of the XYZ \textit{Mandamento}." From such a piece of information we first of all assumed that there exists a link between agents 110 and 77. In addition, we assumed that there exists a link between agent 110 and all the members of the XYZ \textit{Mandamento} (which can be his own or another) identified in the arrest warrant. In Appendix \ref{app:RobustTest} we provide a discussion of this choice and show the results of robustness tests on the topological properties of networks obtained with different hypotheses on the possible number of links between an agent such as agent 110 and the members of different \textit{Mandamenti} in the arrest warrant. The results of the robustness tests are reassuring for the hypothesis we maintain in the main text.} ii) any evidence exists in the arrest warrant of a connection between them. For example: a wire-tapped conversation is reported, a connection is mentioned in a recorded conversation or is simply mentioned in the document; iii) a direct (first-degree) or indirect (second-degree) family tie exists and is documented; iv) the agents participated to a meeting (in this case typically involving more than two persons), documented in the arrest warrant.

This strategy allows to reduce the bias in the identification of the links among the networks' nodes. A large part of the literature on organized crime networks, in fact, relies on information on wiretap data only (e.g. \citealp{varese2012structure}, \citealp{campana2012listening}, \citealp{berlusconi2013all}). However, as pointed out by \citet{berlusconi2013all}, the amount of information in court documents on wiretap data can vary, and in the arrest warrant it is typically lower than in the original set of wiretapped conversations. In addition, as noted by \citet[p. 40]{agreste2015analysis}, individuals at the highest ranks in the organization, i.e. the \textit{bosses}, refrain from using the telephone, knowing that it can be intercepted by the authorities.\footnote{Bernardo Provenzano, the last ``boss of the bosses", was  arrested in 2006 in a small country house not far from his hometown Corleone. Provenzano communicated with the external world through small paper notes (``pizzini"). This communication technique, coupled with diffuse \textit{omerta'} surrounding him, contributed to let him escape police investigations for nearly 40 years. For a concise account of the arrest of Bernardo Provenzano see \citet{butler2007ghost}.} This, obviously, can raise concerns of missing data on both nodes and links of the networks, when the links are identified by wiretap data only. With our strategy this problem is attenuated by exploiting all the information on links among agents in the arrest warrant.

Finally, for the reasons mentioned in Section \ref{secDataSources}, all the detected links will be treated in our empirical analysis as \textit{undirected} and \textit{unweighted}.

\section{Results\label{secResults}}
 In this section we present the results of the empirical analysis of  the \textit{Perseo} network. As remarked, the information on contacts among the individuals involved in the operation allowed us to identify 172 actors, 99 of which were arrested.\footnote{To preserve anonimity, agents have been indexed in the database with a random number between 1 and 172. In the rest of the article the agents will be indicated solely by their number.\label{footRandNum}} The latter group is the main focus of our empirical analysis, as the information in the arrest warrant for this subgroup is the most detailed. Specifically, in Section \ref{secComplexNet} we present the result of the network analysis, while in Section \ref{secEconometrics} we present the results of the econometric analysis based on the estimation of dyadic regressions.

\subsection{Network Analysis\label{secComplexNet}}
All the 99 individuals arrested in the \textit{Operazione Perseo} are from the province of Palermo.\footnote{Two of the 99 agents do not share links with others. In the terminology of social networks, the sub-network of 97 connected agents represents the \textit{giant component} of the \textit{Perseo} network (see, e.g., \citealp[p. 33]{jackson2008social}). As in other studies on networks' topologies \citep[p. 61]{xu2008topology}, in the rest of the paper we focus on the giant component of 97 agents of the \textit{Perseo} network.}  The majority of them is from the city of Palermo (54), while the second largest group (18 members) is from Belmonte Mezzagno, a small town approximately 15 km from Palermo. The agents of the \textit{Perseo} network belong to eight (out of eight) \textit{Mandamenti} of the city of Palermo, and to six (out of seven) \textit{Mandamenti} of the province of Palermo.\footnote{See Figures \ref{fig:diaMandamPa} and \ref{fig:diaMandamPaPr} in Appendix \ref{secAMand} from \citet{dia2016} for the classification and localization of the \textit{Mandamenti}. Each \textit{Mandamento}, especially in the city of Palermo, typically includes different Mafia families. The town of Belmonte Mezzagno belongs to the \textit{Mandamento} of Misilmeri. The distribution of the 97 agents of the \textit{Perseo} network across \textit{Mandamenti} is the following: Brancaccio, Corleone, Partinico, San Mauro Castelverde: 1; Noce, Resuttana:  2; Passo di Rigano - Bocca di Falco, San Lorenzo - Tommaso Natale:  3; Bagheria: 4; Villagrazia - Santa Maria del Gesu': 6; Pagliarelli: 8; Porta Nuova: 16; San Giuseppe Jato: 18; Misilmeri: 20. For eleven agents the \textit{Mandamento} affiliation is not specified in the arrest warrant. \label{foot:MandamDistib}} This suggests that the group under study represents an interesting case for its geographical distribution over a wide area. Following the procedure described in Section \ref{secRecNet}, 470 links among the members of the \textit{Perseo} network have been identified.
 
\begin{figure}[H]
\centering
\includegraphics[scale=0.8,angle=-90]{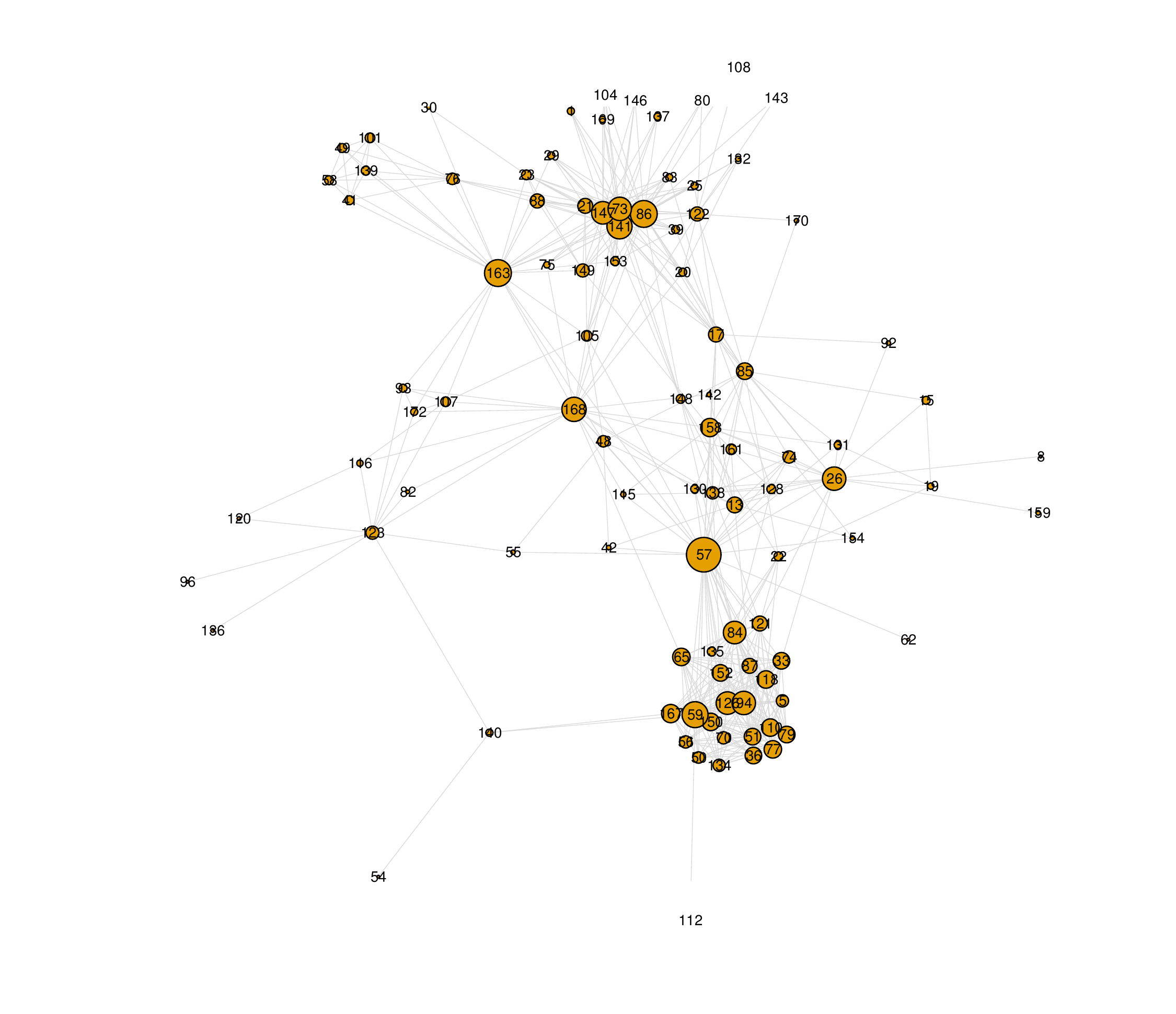}
	
%\vspace{-2cm}
\caption{The \textit{Perseo} network. The size of each node is proportional to the number of its links. The number of each node indicates the identifier each node.}
\label{fig:graphNet99R}
\end{figure}

 Figure \ref{fig:graphNet99R} contains the graphical representation of the \textit{Perseo} network. Figure \ref{fig:graphNet99R} highlights the presence of clusters of densely connected agents, and of some agents (e.g. 168, 86, 57, 163, 26) that appear as \textit{hubs} as they have a relatively high number of links, i.e. a high \textit{degree}.\footnote{Figure \ref{fig:graphNetR} in Appendix \ref{secA172} shows that the larger group of 172 agents mostly includes agents that share some links with agents from the \textit{Perseo} network.} %Some of these agents also appear to play the role of \textit{structural holes}, as they connect otherwise disconnected (or scarcely connected) agents of group of agents \citep{burt1992structural}.

% the list in previous paragraph was: 9,20,27,32,90

The relevant measures characterizing the structure of the network are the following. The density of the network, i.e. the ratio of existing links (470) to the maximum number of possible links, amounts to approximately 10\%.\footnote{The maximum number of links in a network with $N$ agents is $N(N-1)/2$. In the case of $N$=97, this number equals 4656.} In addition, the network has a clustering coefficient of 0.41, a diameter of 5, an average path length of 2.65, and an assortativity level of -0.25.\footnote{The clustering coefficient of a network is the (average) share of links of a node (i.e. who belong to his/her neighborhood) who are also linked. Basically, given a link between agents $i$ and both $j$ and $k$, the clustering coefficient indicates: ``how likely on average is it that $j$ and $k$ are related in the network" \citep[p. 35]{jackson2008social}. The diameter of a network: ``is the largest distance between any two nodes in the network" \citep[p. 33]{jackson2008social}, while the average path length is the average value of the shortest path between any two nodes. The assortativity parameter indicates the tendency of members of the network to be connected to other similar members, where similarity is measured by the their degree (see, e.g., \citealp[p. 65]{jackson2008social}). A relatively high negative number indicates ``disassortativity", i.e. a tendency for individuals with many links to be connected to individuals with few links.} 

How do these topological properties compare to those of other types of networks? Do they represent specific features of a criminal network or they can be found in other networks? To answer these questions Table \ref{tableSNA} compares the values of the measures computed for the \textit{Perseo} network to those characterizing comparable networks: one random network and two small-world networks.\footnote{A random network, in the classic formulation of \citet{erdos1959random} is a network with $n$ agents in which a link between any two agents exists: ``with a given probability $p$, and the formation of links is independent across links" \citep[p. 9]{jackson2008social}. The main characteristic of small-world networks is that they: ``exhibit high clustering and low diameter" \citep[p. 80]{jackson2008social}.} Random networks are unlikely to provide good descriptions of real social networks, but they represent a useful benchmark. Differently, small-world networks may well characterize many real social networks (see the key contribution of \citealp{watts1998collective}).

 \begin{table}[H]
  \centering
 \begin{tabular}{l|ccccc}
  & Density & Clust. Coeff.  & Diameter & Aver. Path Length& Assortativity \\ \hline
\textit{Perseo} Network & 0.10 & 0.41 & 5 & 2.65 & -0.25 \\ 
 Random Network & 0.10 (0.00)  & 0.10 (0.01) & 4 (0.08) & 2.24 (0.01)  & -0.02 (0.04) \\ 
 Small world I & 0.10 (0.00) & 0.41 (0.024) & 4.16 (0.37) & 2.52 (0.04) & -0.01 (0.04)  \\
 Small world II & 0.10 (0.00) & 0.53 (0.02) & 5.06 (0.28) & 2.79 (0.07) & -0.01 (0.04) \\
 \end{tabular}
 \caption{\small{Topological properties of the \textit{Perseo} network, of a comparable random network, and of two comparable small-world networks. The random network and the two small-world networks are simulated 1000 times with the $R$ package \texttt{igraph}. Standard deviations in parenthesis.}}
  \label{tableSNA}
 \end{table}

The values for the random network reported in Table \ref{tableSNA} are the average values from 1000 simulated networks, based on the Erdos-Renyi model \citep{erdos1959random}, with 97 edges and 470 links, i.e. with the same number of nodes and link of the \textit{Perseo} network.\footnote{The random network and the two small-world networks are simulated with the $R$ package \texttt{igraph}.} This ensures that the two networks are comparable, as the same values in Table \ref{tableSNA} for the density of the network confirm. The values reported for Small-world networks I and II are the average values from 1000 simulated networks based on the Watts-Strogatz model \citep{watts1998collective}.\footnote{The algorithm to generate the small-world networks starts from a structure in which each agent is connected to the same number of agents in a regular ring lattice, modified by randomly rewiring some links to agents further away in the lattice (see Figure 1 in \citealp{watts1998collective}). Given this procedure, it was not possible to generate small-world networks with exactly 97 edges and 470 links. However, by considering an initial network of 97 agents in which each agent is initially linked to 5 other agents in the ring lattice, it was possible to generate a network with 485 links, the closest number to 470 (this guarantees that the density is basically the same of the \textit{Perseo} network, see Table \ref{tableSNA}). In order to generate the same (average) clustering coefficient of the \textit{Perseo} network, ``Small-world I" was simualated with a rewiring probability of $0.085$, while in order to generate the same (average) diameter of the \textit{Perseo} network ``Small-world II" was simulated with a rewiring probability of $0.04$.}

In what follows we respectively discuss two main aspects emerging from the social network analysis: the small-world properties of the \textit{Perseo} network, and the centrality of agents in the network.

\subsubsection{The \textit{Perseo} Network is a Small-World Network\label{secSmallWorld}}

Table \ref{tableSNA} suggests that, first of all, the \textit{Perseo} network has a much higher value of the clustering coefficient that in the random counterpart. In addition, the \textit{Perseo} network has a low diameter and average path lengths, although somewhat higher than those of the random network.\footnote{The values of the diameter and of the average path length of the \textit{Perseo} network are in line with the values found by \citet{agreste2016network} for the network of \textit{mafiosi} based on phone calls.} Finally, the \textit{Perseo} network has a much higher level of \textit{disassortativity} with respect to a random network: i.e. agents with many links in the in the \textit{Perseo} network are much more likely to be connected to agents with few links.\footnote{The differences between the values computed from the simulations and those of the \textit{Perseo} network are always highly significant. A simple t-test of the differences between the average values computed from the simulations and the value of the measures from the \textit{Perseo} network always returns p-values equal to zero.} 

A high clustering coefficient and a low diameter (and average path length) imply that the \textit{Perseo} network is a ``small-world" network \citep{watts1998collective}. Previous work identified the small-world property for other criminal networks, such as those studied by \citet{xu2008topology} and \citet{malm2011networks}.\footnote{The criminal networks studied by \citet{xu2008topology} are a jihadist group, a group trafficking in methamphetamine, a group of gangsters and a network of terrorist web-pages, while \citet{malm2011networks} studied a large drug-trafficking network.} To the best of our knowledge, however, no previous study identified the small-world property for the Sicilian Mafia.  Before proposing an explanation for these network properties, we try to answer the following question: are the properties of a small-world network of criminals, given its illicit nature, similar to those of comparable licit small-world networks? To answer this question, the \textit{Perseo} network is compared to two (simulated) small-world networks in Table \ref{tableSNA}: one with the same clustering coefficient (``Small-world I") and one with the same diameter (``Small-world II").

 Table \ref{tableSNA} shows that, with respect to ``Small-world I", the \textit{Perseo} network has a significantly higher diameter. On the other hand, when compared to ``Small-world II"), the clustering coefficient of the \textit{Perseo} network is significantly lower.\footnote{The p-value of a t-test of equality between the two diameter values and the two clustering coefficient values is zero.} 
 
Although the mechanisms explaining small-world networks are fairly well understood,\footnote{In the analysis of social networks, the ``small-world" properties can be explained as follows. A short diameter and average path length (also popularized as the ``six degrees of separation" property) implies that any two members of a large community (e.g. the population of a country) can nevertheless be connected on average by a small number of ``steps". This depends on the fact that within the circle of friends, relatives, acquaintances that any individual has, some contacts have connections that stretch far away from the individual, for example towards citizens of other cities, members of other social groups, etc. The high clustering coefficient, differently, depends on the intuitive fact that, e.g., the friends of an individual with whom they interact frequently, are very likely to get to know each other, forming in this case a cluster of highly-connected individuals.} no compelling explanations exist for this property in criminal networks. Let us recall, first of all, that a criminal network such as the Sicilian Mafia originates within a well-established hierarchical structure, so that the agents' existing links and positions in the network are likely to be determined largely by design and not, at least not exclusively, by the spontaneous activation of isolated individuals.

It is possible to argue that clustering in the \textit{Perseo} network can be explained by the recruitment procedures of \textit{Cosa Nostra}. As pointed out in Section \ref{secIntStruct}, new members are introduced by current members who, having been in the organization for longer time, already have links with others. New members, therefore, are likely to enter what is an already formed cluster, reinforcing this structure as long as they start interacting with the contacts of the member(s) that allowed him to join (see also \citealp[p. 63]{xu2008topology}). However, another crucial element in the recruitment process is that recruitment typically takes place locally. 

The economic theory on criminal organizations, in particular, suggests a specific reason of why this can occur. A criminal organization faces an asymmetric information problem in recruitment which is perhaps even more severe than in legal firms (see in particular \citealp[p. 32]{catino2019mafia}). In fact, criminal organizations must be particularly careful in selecting individuals to recruit not only for a basic problem of unobservability of many of their characteristics, i.e. their perseverance, trustworthiness, etc., but also because the need for secrecy requires that members are not, e.g., undercover officers, that they are not likely to report facts to the police, etc. In this respect, recruiting agents in a geographically close territory is a way to overcome this problem as geographical proximity allows the recruiter to gather information on potential members more easily, for example by directly observing them over time or by collecting information from known sources.\footnote{Indeed, \citet{tumminello2021anagraphical} find that members of the same Mafia syndicate (which does not, however, strictly correspond to a \textit{Mandmanento}), tend to live geographically close.}

\begin{figure}[H]
\hspace{-2cm}
\includegraphics[scale=0.8,angle=-90]{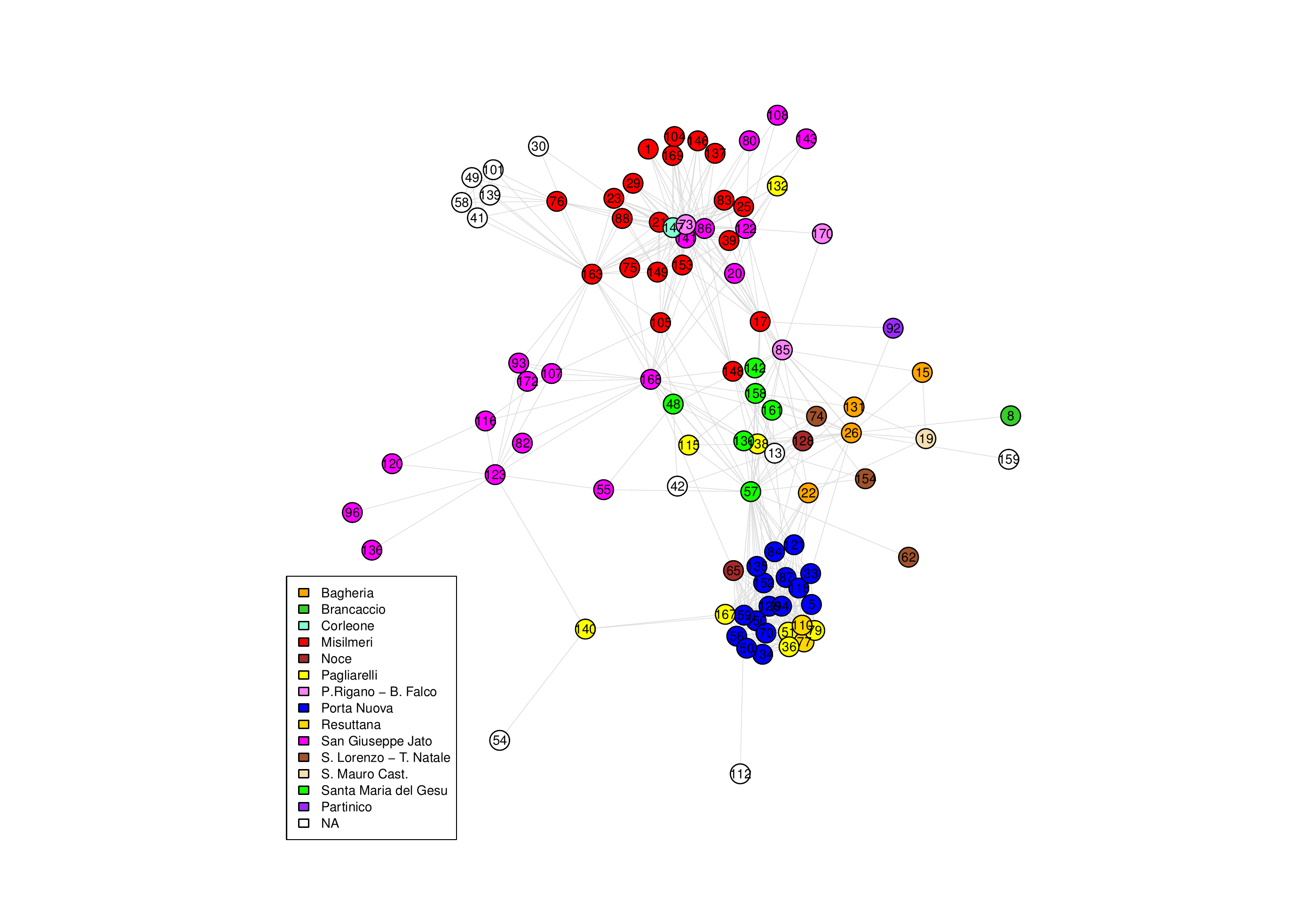}	
	
%\vspace{-2cm}
\caption{The \textit{Perseo} network: Mandamenti highlighted}
\label{fig:graphNet99RCol}
\end{figure}

The rigid territorial organization in \textit{Mandamenti}, moreover, suggests that recruitment by the ruling families is likely to take place within the \textit{Mandamenti} themselves. This implies, in the network representation, that clustering should be high among members of the same \textit{Mandamento}. Figure \ref{fig:graphNet99RCol} suggests that this is the case.

Figure \ref{fig:graphNet99RCol} reproduces the network visualization of Figure \ref{fig:graphNet99R} highlightning the \textit{Mandamento} affiliation of each agent.\footnote{For simplicity, each vertex in Figure \ref{fig:graphNet99RCol} has the same size.} It is apparent that clustering is high among members of the same \textit{Mandamento}, in particular of those of ``Porta Nuova" (blue), ``Misilmeri" (red) and ``San Giuseppe Jato" (magenta).

%\textbf{(FINO QUA)}

The short distance in the network obtains as some members of a highly connected cluster need to connect to agents outside that cluster. The need of organizing the activity over a territory made up of different \textit{Mandamenti} in which in particular illicit trades take place, requires an organizational effort for which some links with individuals geographically further away can improve the efficiency of the organization.\footnote{As pointed out in Footnote \ref{footIntroDoc}, the arrest warrant clearly states that some individuals were explicitly in charge of keeping connections with members of other \textit{Mandamenti}.} Short diameter and average path length, in fact, increase the speed at which information can circulate in the network, making it easier to take timely decisions, organize trades, etc. In particular, given the hierarchically structure of \textit{Cosa Nostra}, the speed of communication can make an order from the higher level reach fast the lower levels.

%The intuition of the economic model of network formation by \citet{jackson2005economics} can be applied here. In the model of \citet{jackson2005economics} agents are located on ``islands". The cost of forming a link among agents on the same island is low, while the cost of forming links among agents belonging to different islands is high. In addition, the benefits of link formation decay with distance among the agents. The implication of this hypothesis is that the resulting efficient network structures feature the small-world properties: high clustering and short distances among agents. 

However, as pointed out before, the level of clustering and the diameter of the \textit{Perseo} network are, respectively, lower and higher that in comparable licit small-world networks. We argue that, along the lines suggested by \citet{baccara2008organize}, this can be explained by the need of a criminal organization to trade off efficiency for secrecy. A somewhat lower level of clustering can indeed reduce the risks faced by the organization as the agents' knowledge of other agents in the same cluster is lower than it would be in an otherwise licit small-world networks.\footnote{For example, in a friendship/acquaintance network,  agents' friends are likely to become friends themselves as nothing in principle prevents an agent to gather together with all of his/her friends. In a Mafia network, on the contrary, an agent can prefer to meet his/her contacts individually, which reduces the probability that they establish connections among themselves.} For the same reason, a higher distance among agents implies that, in terms of secrecy, the agents are more protected as more ``distant" from agents that, if apprehended, can reveal relevant information on other members of the organization. Overall, these results suggest that a Mafia network has properties that correspond to a small-world network, ``adjusted" for its criminal activities.%\footnote{\citet{alizadeh2017generating} study the topological properties of spatial social networks. In particular, they simulate small-world networks in which the probability of link formation exponentially decays with the geographical distance between couples of agents. This type of mechanism, as remarked, can be at work here. \citet[p. 373]{alizadeh2017generating} find that, in a spatial small-world network, the clustering coefficient and the assortativity are \textit{higher} than in a standard small-world network, a result in contrast with our findings. In the econometric analysis of Section \ref{secEconometrics}, however, we do indeed show that the probability of link formation decays with geographical distance. However, it is likely that this effect is dominated in the process of link formation by the secrecy necessities that, we argue, explain the discrepancies in the topological characteristics between the \textit{Perseo} network and comparable small-world.}

Finally, the \textit{Perseo} network displays a high level of ``disassortativity", according to which individuals with many links are more likely to be connected to individuals with few links. This topological characteristic is in stark contrast to what typically characterizes social networks, i.e. a high degree of assortativity \citep[p. 891]{jackson2007meeting}. Indeed, \citet{newman2003mixing} shows that social networks typically display assortative matching, while disassortativity characterize technological and biological networks. Disassortativity, therefore, appears as a peculiar characteristic of criminal networks of Mafia-type. This aspect likely depends on the hierarchical structure, according to which individual at higher levels have many links with individuals at lower levels who are scarcely connected with other members of the organization.\footnote{\citet{xu2008topology} find that disassortativity characterizes the network of methamphetamine network and the dark web network. \citet{xu2008topology} argue that, in the first case, this depends on the hierarchical structure of drug dealing networks, while in the second case on the fact that popular ``dark" web pages receive links from less popular pages.}

%xThese characteristics of the \textit{Perseo} network are in line with the insights from the theoretical analysis of criminal networks of \cite{baccara2008organize}: in a criminal organization we should expect some parsimony in the creation of links, given that establishing a new link has a cost in terms of increased risks for the organization. On the other hand, an efficient communication structure is essential for the functioning of the organization, especially in the case of a Mafia network, which is engaged in the daily activities related to the need of controlling the territory, such as running an extortion racket, smuggling drugs, and providing different sorts of protection services (see \citealp{lavezzi2014organised}, for details and references). In general, these results suggest that a Mafia network might possess properties that correspond to a small-world network, ``adjusted" for its criminal activities.

\subsubsection{Mafia Bosses and Network Centrality\label{secBossesCentr}}

In this section the focus is on the agents that, according to the investigation, had a directive role in the organization. For short, they will be called ``bosses". From the available evidence, 14 agents in the \textit{Perseo} network were bosses. As discussed in Section \ref{secIntStruct}, the Sicilian Mafia has a hierarchical structure, so some \textit{mafiosi} are at higher levels than sub-groups of other members that they command. Understanding leadership in criminal organizations is not only interesting for a better knowledge of these groups, but can also inform policies aimed at contrasting organized crime, as apprehension of leaders can be decisive to weaken the organization. In this section we will analyze how the position in the hierarchical structure is related to the position in the network, an issue that received a great deal of attention by scholars (see, among others, \citealp{borgatti2006identifying} and \citealp{calderoni2019nature}).

In social network analysis, the relevance of an agent in the network is evaluated through various measures of \textit{centrality} (see, e.g. \citealp[pp. 37-43]{jackson2008social}). The most intuitive measure of centrality is denoted as \textit{degree centrality} and is measured by the number of links of an agent. Such number is denoted for short as the \textit{degree} of an agent. Although different measures of centrality can characterize the different roles that agents have within a network,\footnote{See, among others, the discussions in \citet[pp. 37-43]{jackson2008social}, \citet{calderoni2019nature}, \citet{borgatti2006identifying}. In \citet{battistiEtAlSpatial2022} we provide a spatial econometric analysis of centrality in the \textit{Perseo} network. Our main finding is that there exist positive spatial effects among the eigenvector centrality scores. An agent has a high eigenvector centrality score if s/he is connected to agents with a large degree. The  result suggests that agents with high eigenvector centrality tend to be located at short distance in the geographical space. See \citet{battistiEtAlSpatial2022} for more details and a discussion of policy implications.} \textit{degree centrality} is straightforwardly connected to the visibility of an agent and to the security of the organization. As pointed out by \citet{baccara2008organize}, in fact, having links with other agents exposes a member of the organization to the risk of being identified and punished by an external authority if one of his contacts is identified and reveals information on other members.

\begin{figure}[H]
\centering
\includegraphics[scale=0.5, angle=-90]{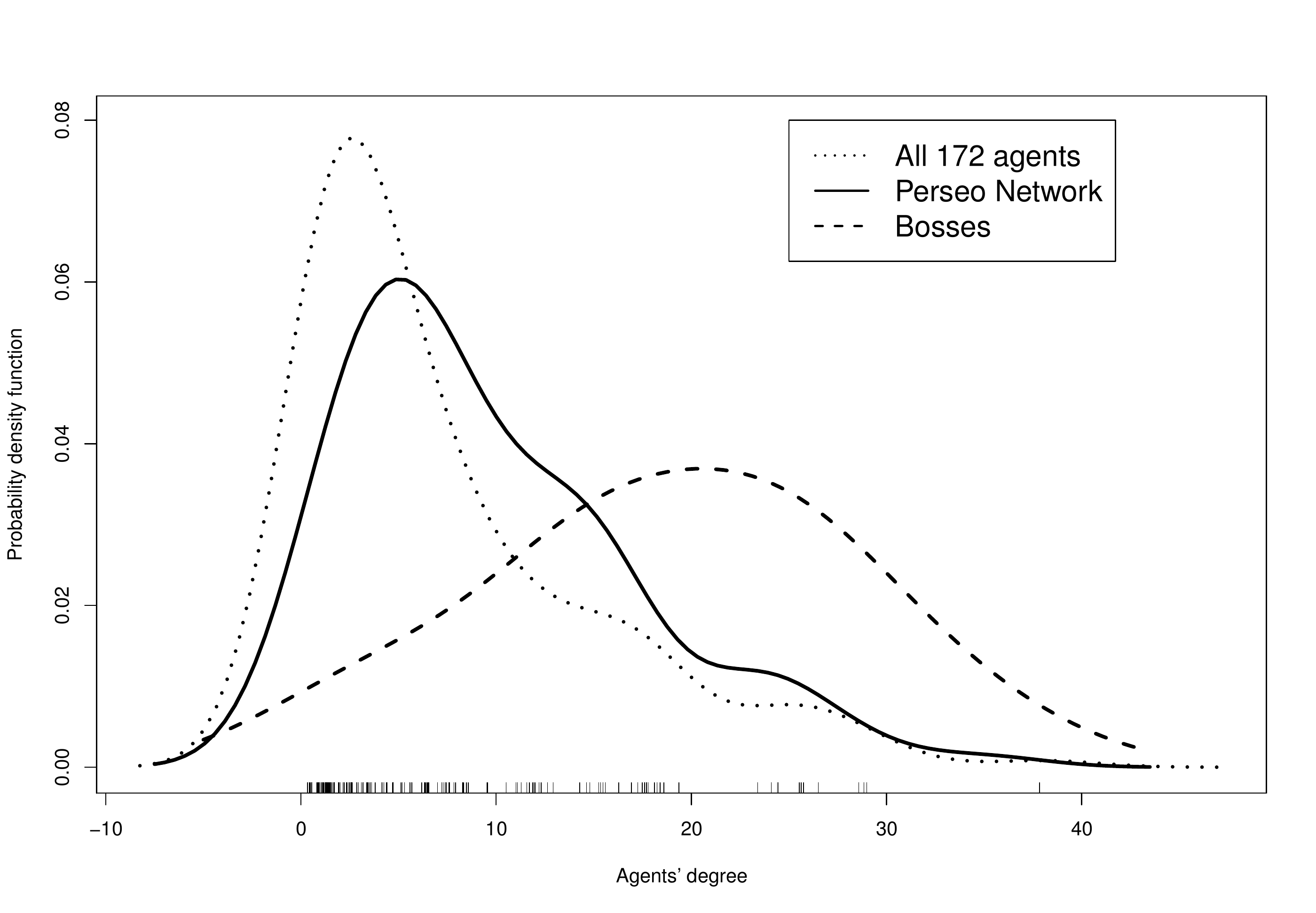}	
\caption{Degree distributions of: i) the whole set of 172 agents; ii) the core group of 99 agents; iii) bosses}
\label{fig:degreeDens}
\end{figure}

Figure \ref{fig:degreeDens} presents an estimation of the degree distribution of the \textit{Perseo} network. In particular, Figure \ref{fig:degreeDens} compares the degree distribution of the \textit{Perseo} network to the degree distribution of the larger set of 172 agents identified in the arrest warrant, and to the degree distribution of the subset of bosses.\footnote{The densities in Figure \ref{fig:degreeDens} are estimated non-parametrically, with normal optimal smoothing (see \citealp[p. 31,]{bowman1997applied} for details). For better visual representation, the estimation assumes that some degree levels could be negative.}

Figure \ref{fig:degreeDens} highlights that the degree distribution of of the \textit{Perseo} network is more skewed to the right than the degree distribution of the network with 172 agents, consistent with the fact that the larger network includes more agents with a relatively lower number of links, as these agents are typically connected to one or few member of the core group (see Figure \ref{fig:graphNetR} in Appendix \ref{secA172}).\footnote{In Appendix \ref{app:ScaleFree} we show that, however, the \textit{Perseo} network is not scale-free. Scale-free networks are characterized by a strongly asymmetric distribution of degrees.} The subset of bosses has a degree distribution even more skewed to the right. This points out that, on average, they have more contacts than the network population at large. This is consistent with the existence of a hierarchical structure within the organization: the bosses rule over a certain number of other members which implies, from the social network perspective, a higher number of links than those at a lower hierarchical level.

\begin{figure}[H]
\hspace{-2cm}
\includegraphics[scale=0.9,angle=-90]{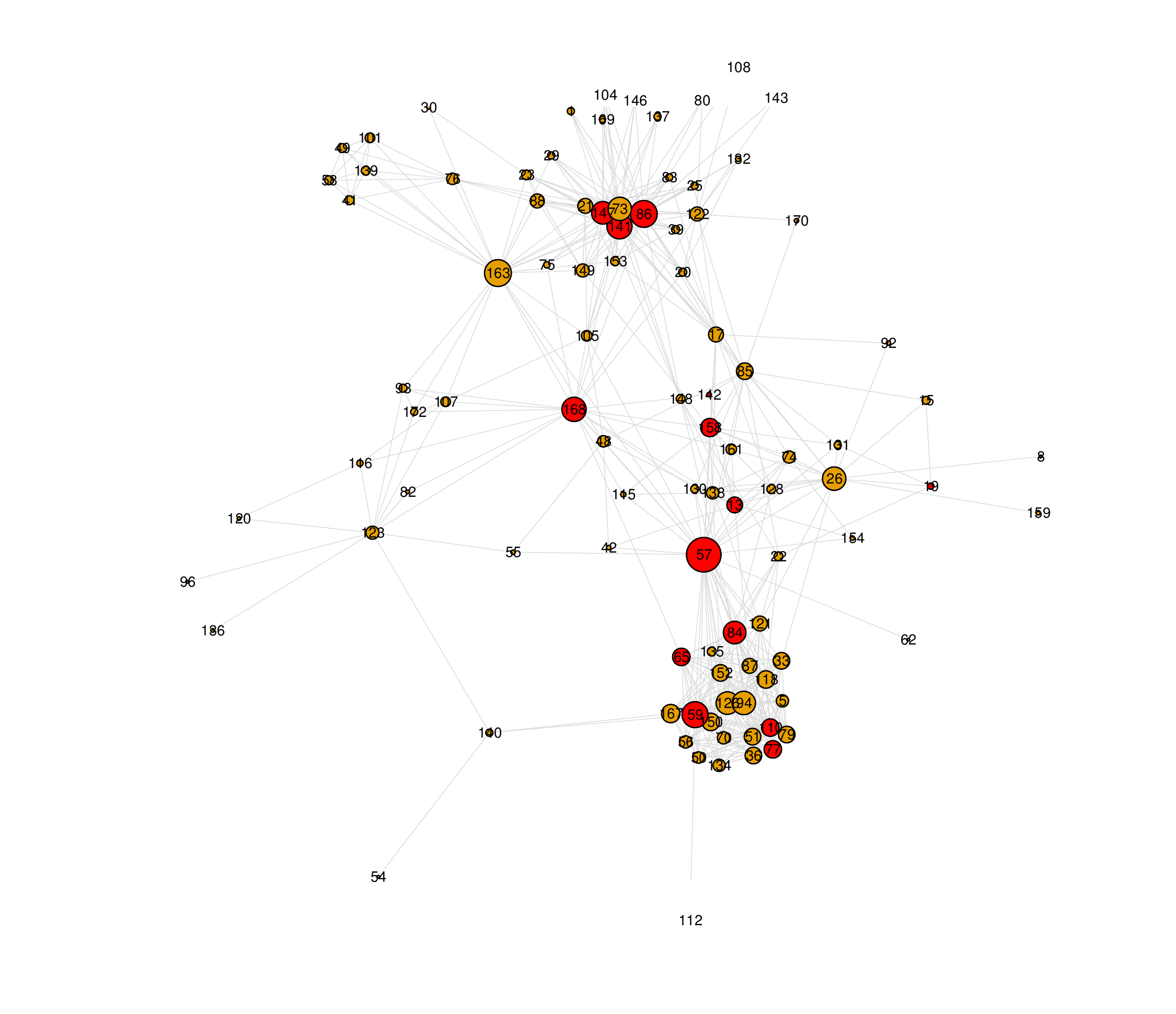}		

%\vspace{-2cm}
\caption{The \textit{Perseo} network: agents with directive roles highlighted}
\label{fig:graphNet99RD}
\end{figure}

However, the correlation between being a boss and having a high degree is not perfect, as Figure \ref{fig:graphNet99RD} shows. Figure \ref{fig:graphNet99RD} is a reproduction of Figure \ref{fig:graphNet99R} in which the bosses are highlighted in red. Interestingly, although in many cases bosses have high degrees, there are agents with high degrees that are not bosses (e.g. 26, 94, 163), and bosses with low degrees (e.g. 13, 10, 65, 77). 

To shed more light on this issue, Table \ref{tab:degreeDir} ranks the agents with the 10 highest degrees, reporting the age of the agent and the indication of family ties within the organization. In fact, longevity in the organization (proxied by age) or a family tie, can be good predictors of having a directive role in the organization (see, e.g., \citealp[p. 42]{paoli2004italian}). Table \ref{tab:degreeDir} suggests that, indeed, the connection between degree centrality and being a boss is more nuanced when one takes into account the age of the agent and having or not family ties with someone else in the organization.

\begin{table}[H]
\centering
\begin{tabular}{ccccc}  \hline

Id & Degree & Boss & Age & Family ties \\ 
  \hline
  57 & 35 & YES & 31 & YES \\ 
86 & 29 & YES & 43 & NO \\ 
  141 & 27 & YES & 73 & NO \\ 
    59 & 25 & YES & 84 & YES \\ 
  73 & 25 & NO & 45 & NO \\ 
    163 & 24 & NO & 56 & NO \\ 
    147 & 24 & YES & 59 & NO \\ 
    94 & 23 & NO & 62 & NO \\ 
   84 & 23 & YES & 56 & NO \\ 
     168 & 22 & YES & 62 & NO \\ 

\end{tabular}
\caption{The 10 actors with the highest degrees in the \textit{Perseo} network. ``Direction" indicates whether investigators attributed a directive role to the actor. ``Age" is the age of the agent in 2008, ``family ties" indicates whether the agent has family ties with other members of the \textit{Perseo} network}
\label{tab:degreeDir}
\end{table}

The agent with the highest degree, in particular, is of relatively young age, but happens to have a family tie with other members of the organization. In particular, he is related to an important figure in the network.\footnote{To preserve anonymity we remain vague on the exact information on these two agents.}

\begin{figure}[H]
\hspace{1cm}
\begin{minipage}[t]{0.40\linewidth}
\includegraphics[width=6cm]{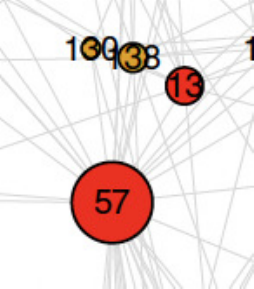}

  \caption{Detail from Figure  \ref{fig:graphNet99RD}.}
  \label{fig:netZoom}
    \end{minipage}
\begin{minipage}[t]{0.40\linewidth}
	\includegraphics[width=5.5cm]{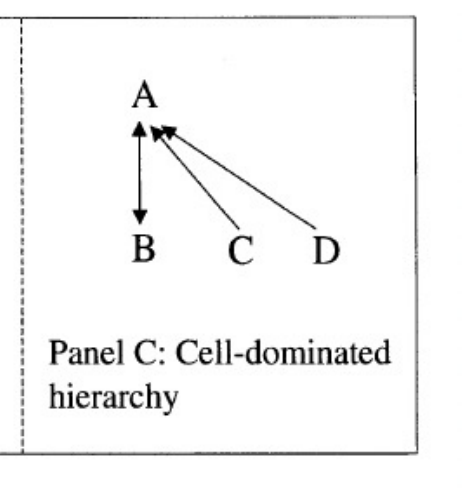}	
	
  \caption{Detail from Figure 2 of \citet{baccara2008organize}.}
  \label{fig:cellDomHierBaccara}

\end{minipage}
\end{figure}

An interesting aspect of these findings is that a crucial aspect of the configuration of the \textit{Perseo} network is consistent with a key finding of \citet{baccara2008organize}. Figures \ref{fig:netZoom} and \ref{fig:cellDomHierBaccara} show that we can retrace a case of ``cell-dominated hierarchy" \citep[p. 1049]{baccara2008organize} in the \textit{Perseo} network.

A ``cell-dominated hierarchy" is a network configuration in which the information hub, i.e. the agent holding information on other agents, is part of a binary cell. In Figure \ref{fig:cellDomHierBaccara}, $A$ and $B$ represent a binary cell. Assuming that $B$ is an important member or the organization, this is efficient because agents $C$ and $D$ in Figure \ref{fig:cellDomHierBaccara} are insulated from agent $B$ and, therefore, if apprehended they cannot reveal information on him. As formally demonstrated by \citet{baccara2008organize} this maximizes the efficiency of the network, as the information hub can contact all agents, but it does so guaranteeing higher security to the network, as information on agent $B$ could not revealed by agents $C$ and $D$.\footnote{Differently, with links between $B$ and the couple of agents ($C$,$D$), the same circulation of information could be attained, but with lower security.}

The detail of the \textit{Perseo} network in Figure \ref{fig:netZoom} suggests that agents 57 and 13 could be considered as a binary cell. Agent 13 has a higher position in the organization, as the court evidence demonstrates. It is not him, however, that keeps a high number of contacts with other members, but an agent connected to him, agent 57 a relative. From the security point of view, this makes sense: the high-level individual is not central in the network, and delegates a trusted person (a relative) to keep the contacts with the rest of the organization. As pointed out elsewhere \citep{agreste2016network,calderoni2019nature}, network centrality does not always imply leadership in the organization. We add to this that the possibility that the more central agents can be found in the family of the bosses. This pieces of information are important for the design of policies to disrupt criminal organizations by apprehending key members.

In the next section we present the results of the econometric analysis of the \textit{Perseo} network, that will shed more light on the its structure, exploiting in particular information on the members of the organization.

\subsection{Econometric Analysis\label{secEconometrics}}
Although the network analysis is informative on key aspects of the organizational structure, it does not allow to conduct a more rigorous statistical analysis of the existing links among the agents, in particular with respect to the available information on individuals' characteristics (i.e. age, task within the organization, etc.). In this section, therefore, we present the results of the econometric analysis of link formation among members of the \textit{Perseo} network, as complementary to the social network analysis of Section \ref{secComplexNet}.

Specifically, we aim at identifying factors that display robust correlations with the detected links. Following \citet{fafchamps2007formation}, we estimate dyadic regressions in which  two types of variables are considered: i) variables representing attributes of the link between agents $i$ and $j$, denoted as $w_{ij}$; ii) variables representing individual attributes of agent $i$, denoted as $z_i$, that might correlate with the links that $i$ has with other agents.

The aim of this analysis is answering questions such as the following: does the probability that a link exists between $i$ and $j$ decrease or increase with the geographical distance between them? Does it decrease or increase with the difference in age, i.e. a measure of social distance? Are individuals with the same task in the organization more or less likely to be connected? These questions are considered in turn.

The dyadic regressions, in the case of the estimation of correlates of the attributes of a link between $i$ and $j$, are estimated as a logit model where the dependent variable is a binary variable taking value 1 if a link between agents $i$ and $j$ exists or 0 otherwise, i.e. if, denoting by $\mathbf{A}$ the adjacency matrix representing the \textit{Perseo} network,\footnote{Given a network among $N$ agents, the adjacency matrix $\mathbf{A}$ is a $N\times N$ matrix in which $A_{ij}=1$ if a link between $i$ and $j$ exists and 0 otherwise. If the links are undirected, as in our case, then the matrix is symmetric and $A_{ij}=A_{ji}$. Since a link from $i$ to $i$ is not defined, $A_{ii} = 0, \forall i$ by construction.} the element $A_{ij}=A_{ji}=1$.  The relationship between the set of covariates $\mathbf{X}_{ij}$ and the probability that $A_{ij}=A_{ji}=1$, i.e. that a link between $i$ and $j$ exists, is therefore described by the following logistic equation:

\begin{equation}
P(A_{ij}=A_{ji}=1|\mathbf{X}_{ij})=\frac{exp(\beta_{0}+\beta _{1}\mathbf{X}_{ij}+\epsilon_{ij})}{1+exp(\beta _{0}+\beta _{1}\mathbf{X}_{ij}+\epsilon_{ij})}
\label{dyadReg}
\end{equation}

\vspace{4mm}

 Eq. (\ref{dyadReg}) is estimated with respect to indicators of links' attributes, $w_{ij}$, and separately with respect to individual's characteristics $z_i$.\footnote{In the case of estimations based on the attributes of agent $i$, the formulation of Eq. (\ref{dyadReg}) should be accordingly modified.} We run two types of regressions because, differently from \citet{fafchamps2007formation}, the links of the \textit{Perseo} network are undirected. For this reason,  only the information related to the upper (or lower) triangular part of the adjacency matrix $\mathbf{A}$ is utilized to estimate the effect of links' attributes $w_{ij}$, while the information related to all links is considered when estimating the effects of individual's characteristics $z_i$ on link formation.\footnote{In the study of \citet{fafchamps2007formation} on the formation of risk-sharing networks, agent $i$ can mention agent $j$ as a contact while the opposite might not occur. For this reason, in their adjacency matrix $\mathbf{L}$, $L_{ij} =1$ does not imply $L_{ji}=1$. In this case, therefore, in the estimation they use all the links appearing in $\mathbf{L}$ and consider in the same regression both $w_{ij}$ and $z_i$. In our case, differently, when estimating the effect of $w_{ij}$, considering both elements $A_{ij}$ and $A_{ji}$ would introduce redundant information in the estimation, as for any $w_{ij}$ related to $A_{ij}$, there would appear the same $w_{ij}$ related to $A_{ji}$.}

The links' attributes $w_{ij}$ we consider are: i) the Euclidean distance between $i$ and $j$, denoted as $d_{ij}$, computed from the geo-localized (by latitude and longitude) addresses of the agents. We also control for possible nonlinear effects of $d_{ij}$; ii) the age difference between $i$ and $j$, denoted as $Age_{|i-j|}$; iii) a variable indicating the number of tasks in common between $i$ and $j$, denoted as $f_{i \equiv j}$. We identified eight tasks (see Section \ref{secData}): the variable $f_{i \equiv j}$ takes the values $0-5$ as, at most, some agents shared $5$ functions; iv) a dummy taking the value $1$ if both $i$ and $j$ are bosses, denoted as $Boss_{i \equiv j}$; v) a dummy taking the value $1$ if both $i$ and $j$ belong to the same \textit{Mandamento}, denoted as $M_{i \equiv j}$; vi) two variables indicating the (absolute value) of the difference between two indicators of network's centrality, the degree and the value of the eigenvector centrality index, as another possible relevant measure of centrality, denoted respectively as $DegCent_{|i-j|}$ and $EigCent_{|i-j|}$.\footnote{The \textit{eigenvector centrality} is high when an agent is connected to highly-connected agents (see \citealp[p. 39]{jackson2008social}).}Table \ref{tab:resDyads01newTab4} contains the results.\footnote{Since observations in dyadic regressions are not independent, we consider bootstrapped standard errors with 200 replications. Results are not affected by increasing the number of replications to 500.}

\begin{table}[H]
\centering
\scriptsize
{
\def\sym#1{\ifmmode^{#1}\else\(^{#1}\)\fi}
\begin{tabular}{l*{8}{c}}
\toprule
                    &\multicolumn{1}{c}{(1)}&\multicolumn{1}{c}{(2)}&\multicolumn{1}{c}{(3)}&\multicolumn{1}{c}{(4)}&\multicolumn{1}{c}{(5)}&\multicolumn{1}{c}{(6)}&\multicolumn{1}{c}{(7)}&\multicolumn{1}{c}{(8)}\\
%                    &\multicolumn{1}{c}{link}&\multicolumn{1}{c}{link}&\multicolumn{1}{c}{link}&\multicolumn{1}{c}{link}&\multicolumn{1}{c}{link}&\multicolumn{1}{c}{link}&\multicolumn{1}{c}{link}&\multicolumn{1}{c}{link}\\
\midrule
%link                &               &               &               &               &               &               &               &               \\
$d_{i,j}$                &      -2.543***&      -4.314***&      -4.419***&      -4.306***&      -4.556***&      -1.367   &      -1.638*  &      -0.646   \\
                    &     (0.714)   &     (1.134)   &     (1.023)   &     (1.036)   &     (1.055)   &     (1.067)   &     (0.974)   &     (1.084)   \\
\addlinespace
$d_{i,j}^2$  &               &       3.194** &       3.329***&       2.925** &       2.402** &      -1.024   &      -0.750   &      -1.514   \\
                    &               &     (1.241)   &     (1.146)   &     (1.198)   &     (1.214)   &     (1.428)   &     (1.322)   &     (1.401)   \\
\addlinespace
$Age_{|i-j|}$            &               &               &       0.005   &       0.000   &      -0.008*  &      -0.006   &      -0.006   &      -0.009*  \\
                    &               &               &     (0.005)   &     (0.004)   &     (0.004)   &     (0.004)   &     (0.004)   &     (0.004)   \\
\addlinespace
$F_{i \equiv j}$     &               &               &               &       0.428***&       0.307***&       0.267***&       0.260***&       0.221***\\
                    &               &               &               &     (0.066)   &     (0.068)   &     (0.074)   &     (0.070)   &     (0.078)   \\
\addlinespace
$Boss_{i \equiv j}$      &               &               &               &               &      -1.182***&      -1.421***&      -1.391***&      -1.339***\\
                    &               &               &               &               &     (0.103)   &     (0.121)   &     (0.118)   &     (0.120)   \\
\addlinespace
$M_{i \equiv j}$          &               &               &               &               &               &       1.631***&       1.635***&       1.667***\\
                    &               &               &               &               &               &     (0.146)   &     (0.139)   &     (0.139)   \\
\addlinespace
$DegCent_{|i-j|}$         &               &               &               &               &               &               &       0.019   &       0.012   \\
                    &               &               &               &               &               &               &     (0.012)   &     (0.012)   \\
\addlinespace
$EigCent_{|i-j|}$      &               &               &               &               &               &               &               &       1.411***\\
                    &               &               &               &               &               &               &               &     (0.210)   \\
\addlinespace
Constant            &      -1.892***&      -1.778***&      -1.850***&      -2.138***&      -1.089***&      -1.449***&      -1.573***&      -1.906***\\
                    &     (0.084)   &     (0.104)   &     (0.104)   &     (0.121)   &     (0.154)   &     (0.168)   &     (0.179)   &     (0.178)   \\
\midrule
N                   &    4560   &    4560   &    4560   &    4560   &    4560   &    4560   &    4560  &    4560   \\
aic                 &    2983.556   &    2980.317   &    2980.755   &    2935.387   &    2818.257   &    2698.957   &    2697.607   &    2656.215   \\
bic                 &    2996.406   &    2999.592   &    3006.455   &    2967.512   &    2856.807   &    2743.933   &    2749.008   &    2714.041   \\
\bottomrule
\multicolumn{9}{l}{\footnotesize Standard errors in parentheses}\\
\multicolumn{9}{l}{\footnotesize * p<0.10, ** p<0.05, *** p<0.010}\\
\end{tabular}
}

\caption{Dependent variable=1 if $i$ and $j$ are linked. Estimation is logit, standard errors are bootstrapped with 200 replications. * p$<$0.10, ** p$<$0.05, *** p$<$0.010}
\label{tab:resDyads01newTab4}	
\end{table}

%\begin{table}[H]
%\scriptsize
%\centering
%\input{resDyads01copia.tex}
%\caption{Dependent variable=1 if $i$ and $j$ are linked. Estimation is logit, standard errors are bootstrapped with 200 replications. * p$<$0.10, ** p$<$0.05, *** p$<$0.010}	
%\label{tabDyad}
%\end{table}

\begin{figure}[H]
\centering
\includegraphics[angle=-90, scale=0.3]{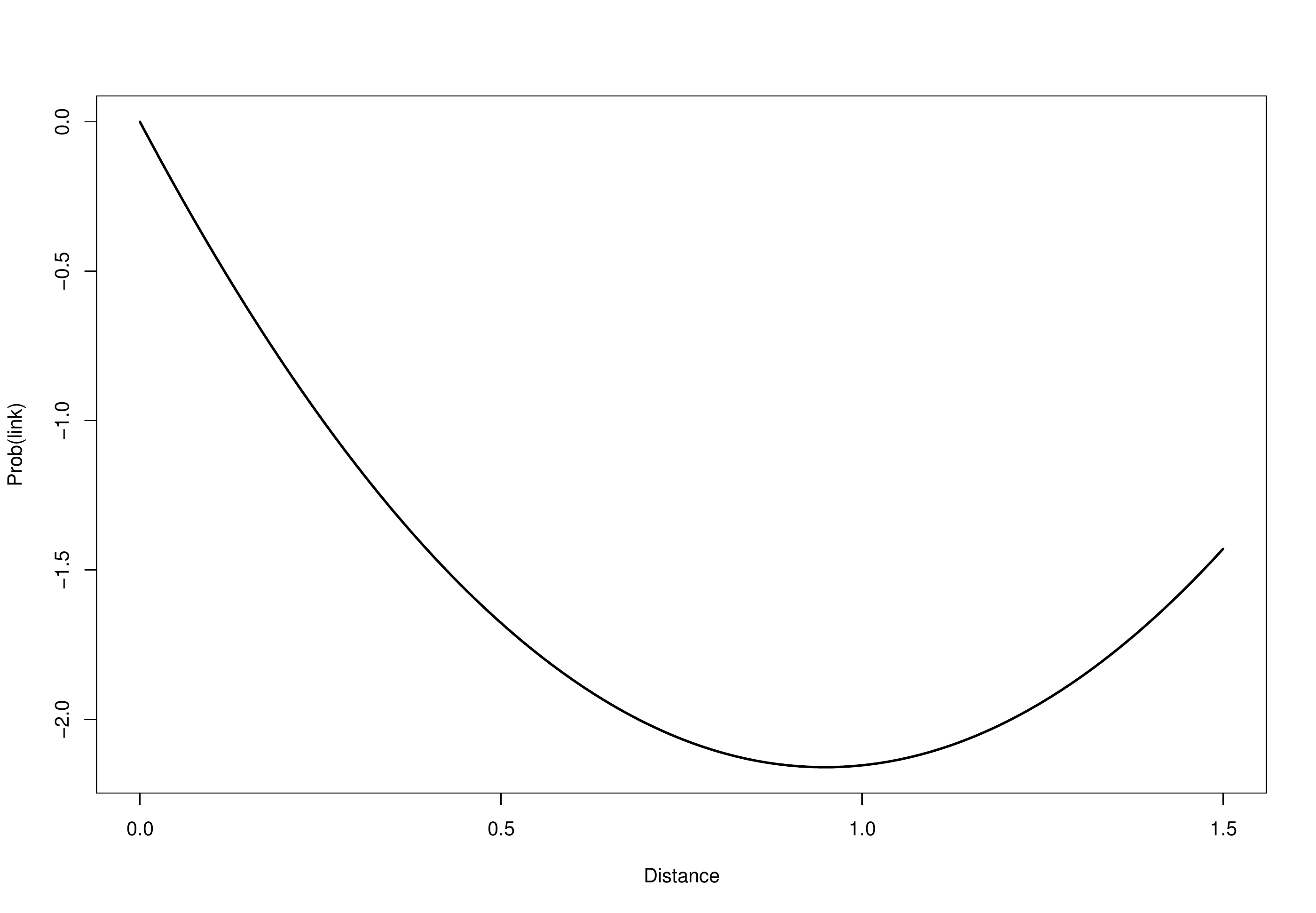}

\caption{Relationship between geographical distance and link formation as from Model (5) of Table \ref{tab:resDyads01newTab4}}
\label{fig:plotDist}
\end{figure}

Models (1) and (2) of Table \ref{tab:resDyads01newTab4} show that the geographical distance between agents $i$ and $j$ is significantly correlated with the existence of a link, displaying a robust nonlinear effect.\footnote{If we take the addresses of the agents ad exogenous, and pre-determined with respect to the existence of a link, than this effect could be considered as causal. Since, however, we cannot control for the address of an agent at birth, it is not possible to rule out the event that the choice of where to reside depends on the social contacts of an agent. Unfortunately, data does not allow to control for this, although the latter possibility seems quite unlikely.} As Figure \ref{fig:plotDist} shows, the probability of link formation decreases with distance up to a point in which it starts increasing.\footnote{This figure represents the marginal effects of distance with respect to the probability to be linked and is plotted with the coefficients of Model (5), which considers other covariates.} This results is partially in line with the theory of spatial social networks (e.g. \citealp{alizadeh2017generating}), which suggests that the probability that two agents are connected, decreases with the geographical distance between them. However, from a certain point on, we find that the probability increases. This suggests that above some threshold distance, it becomes optimal again to establish links with an another agent, located further away.

Model (3) considers the effect of difference in age which is, however, non significant. This suggests that neither the difference nor the similarity in age significantly contributes to the probability of link formation (the coefficient for $Age_{|i-j|}$ is basically non significant in the other specifications). Model (4) shows that the variable indicating the similarity in the number of tasks between agents has a positive and highly significant effect: the more ``similar" the agents are in terms of the functions they perform, the more likely they are linked. This suggests that the work organization within the network implies higher probability of interaction (which can imply collaboration, sharing of information, etc.) among agents who are more similar to each other in terms of the tasks they perform, an aspect that is likely to increase the efficiency of the organization itself. As pointed out by works such as \citet{kleinbaum2013discretion}, the organizational needs can put constraints on the patterns of interactions within the members of the organization, which seems to be the case here. To the best of our knowledge the correlation between task specialization and the structure of interaction in a criminal organization was not noted before.%\footnote{\citet{campana2012listening} and \citet{varese2012mafias} discuss the use of content analysis and correspondence analysis in the study of criminal organizations. By analazyng conversations from court evidence these techniques allow for the classification of agents according to their tasks and the association of the tasks performed with some indices of homophily, for example ethnicity or nationality. These techniques, however, do not allow for the estimation of homophily indices on link formation.}

 Model (5) shows that having a directive role has a significant \textit{negative} effect on the probability of link formation between two agents. This suggests that two agents with directive roles are not likely to be connected, in accordance with the results of Section \ref{secComplexNet}, pointing out the hierarchical structure of the organization and the need for secrecy, so individuals with directive roles (and perhaps relevant information about the organization) are less likely to be connected than with agents without directive roles. Model (6) shows that belonging to the same \textit{Mandamento} has a strong and significant positive effect on link formation and, in particular, it makes the effect of geographical distance non significant. This suggests that the size of the \textit{Mandamenti}  might actually be optimal in the sense that it (roughly) corresponds to the distance level that minimizes the probability of link formation. Alternatively, this may suggests that the joint membership to a \textit{Mandamento}, which is also related to the geographical distance among agents, is what matters for the creation of links as, e.g., it may guarantee that an agent is trustworthy. In this respect, joint membership to a \textit{Mandamento} might be comparable to the role of ethnicity or nationality in defining homophily between agents, that can increase the probability that they are connected in a social network (see, e.g, \citealp{bouchard2020collaboration}, for discussion and references). Finally, Model (7) shows that the difference between in degree centrality is not significant, while the difference in eigenvector centrality is positive and highly significant. The latter results point out that the larger the difference in eigenvector centrality, the higher the probability that a link exists between two agents, another aspect of the high disassortativity in the \textit{Perseo} network.

Table \ref{tab:resDyadsFull02newTab5} contains the results of the estimation of Eq. (\ref{dyadReg}) when the set of covariates $\mathbf{X}_{ij}$ only contains the individual characteristics $z_i$. Now the estimated coefficients refer to the correlation between an individual characteristic and the probability that a link emanates from an agent with a certain level of that characteristic (or with that characteristic in the case of dummies). The individual characteristics that we consider are: i) agent's age (\textit{Age}); ii) the number of tasks performed by an agent (a number comprised between 1 and 5, denoted by $N_k$); iii) dummy variables related to individual tasks: drug dealing (\textit{Drugs}), smuggling guns (\textit{Guns}), taking care of connections within the \textit{Mandamento} ($L_{IN}$), taking care of connections outside the \textit{Mandamento} ($L_{OUT}$), collecting extortion money (\textit{Pizzo}), dealing with public procurement (\textit{PubProc}), organizing meeting (\textit{Meeting}), being a boss (\textit{Boss});\footnote{In particular, the dummy variable \textit{Boss} summarizes the different ranks that, according to the investigators, a member had in the organization: family leader or \textit{Mandamento} leader (\textit{Boss}=1), otherwise (\textit{Boss} = 0).} iv) a dummy for having family ties with other members (\textit{Family}); v) \textit{Mandamento} fixed effects, to control for the possible effect on links formation given by organizational features that might be related to belonging to the family(ies) ruling on some \textit{Mandamento}.

Model (1) contains \textit{Age} as the only covariate. Age can increase the probability that an agent has links with others, if being older implies being connected to many agents for the longevity in the organization. Model (1) shows that, indeed, age has a highly significant positive effect on the probability of link formation.

Model (2) adds $N_k$. This variable has a highly significant correlation with the probability of link formation of an agent: the higher the number of tasks, the higher the probability the agent has links with other agents. \textit{Age} is still positive and significant.

In Model (3) we  consider, instead of $N_k$, all the dummy variables referred to the different tasks identified by the investigators, to try to disentangle the possible effects of the distinct task. We can observe that $L_{OUT}$ and \textit{Pizzo} have positive, highly significant coefficients, while \textit{PubProc} has a positive coefficient, significant at 10$\%$ confidence level. This suggests that having the task of establishing connections outside the \textit{Mandamento} or dealing with the collection of protection money are particularly relevant for the probability of link formation.

In Model (4) we add the dummy \textit{Boss}. It is highly significant, pointing out that being in charge of directing the organization implies having a high probability of link formation with others. In addition, $L_{OUT}$ and \textit{pizzo} are still highly significant, although the estimated coefficient for $L_{OUT}$ is much lower, while \textit{PubProc} loses its significance. This suggests that $L_{OUT}$ and \textit{PubProc} alone picked up part of the effect that, in fact, should be attributed to being a boss.\footnote{Interestingly, Table \ref{corrtable} in Appendix \ref{AppOtherRes} highlights that, in fact, the tasks with the highest correlations with a being a boss are \textit{LinkingOut} and \textit{PubProc} suggesting that those in charge of directing part of the organization take care of delicate tasks such as establishing links outside the \textit{Mandamento} or dealing with public procurement, which implies interacting with the political power.} Finally, the coefficient for \textit{Age} is halved, and loses significance.\footnote{The correlation between \textit{Age} and \textit{Boss} is 0.24.}

In Model (5) a dummy variable for having family ties with others agents in the \textit{Perseo} network is added, but this variable has a nonsignificant effect. Finally, in Model (6) we introduce \textit{Mandamento} fixed effects. The coefficients for $L_{OUT}$, and \textit{Boss} remain positive and highly significant, in line with the previous models. Differently, the coefficient of \textit{Guns} has the same sign but its absolute value and significance strongly increase, \textit{Pizzo} loses significance, while   \textit{PubProc} becomes significant with a negative sign. We take the results of Model (6) with some caution as the number of observations is strongly reduced with respect to Models (1)-(5), as a for a certain number of agents the indication of \textit{Mandamento} is not available (see Footnote \ref{foot:MandamDistib}).

\begin{table}[H]
\centering
\scriptsize
{
\def\sym#1{\ifmmode^{#1}\else\(^{#1}\)\fi}
\begin{tabular}{l*{6}{c}}
\toprule
                    &\multicolumn{1}{c}{(1)}&\multicolumn{1}{c}{(2)}&\multicolumn{1}{c}{(3)}&\multicolumn{1}{c}{(4)}&\multicolumn{1}{c}{(5)}&\multicolumn{1}{c}{(6)}\\
%                    &\multicolumn{1}{c}{"link"}&\multicolumn{1}{c}{"link"}&\multicolumn{1}{c}{"link"}&\multicolumn{1}{c}{"link"}&\multicolumn{1}{c}{"link"}&\multicolumn{1}{c}{"link"}\\
\midrule
%"link"              &               &               &               &               &               &               \\
\textit{Age}               &       0.015***&       0.012***&       0.010***&       0.005*  &       0.006** &      -0.000   \\
                    &     (0.003)   &     (0.002)   &     (0.002)   &     (0.003)   &     (0.003)   &     (0.003)   \\
%\addlinespace
$N_k$            &               &       0.247***&               &               &               &               \\
                    &               &     (0.032)   &               &               &               &               \\
%\addlinespace
\textit{Drugs}           &               &               &       0.112   &      -0.062   &      -0.105   &       0.187   \\
                    &               &               &     (0.181)   &     (0.196)   &     (0.206)   &     (0.403)   \\
%\addlinespace
\textit{Guns}            &               &               &       0.230*  &       0.272*  &       0.257*  &       0.662***\\
                    &               &               &     (0.131)   &     (0.156)   &     (0.146)   &     (0.203)   \\
%\addlinespace
$L_{IN}$         &               &               &       0.002   &      -0.016   &      -0.018   &      -0.007   \\
                    &               &               &     (0.099)   &     (0.101)   &     (0.101)   &     (0.141)   \\
%\addlinespace
$L_{OUT}$        &               &               &       0.587***&       0.395***&       0.390***&       0.576***\\
                    &               &               &     (0.095)   &     (0.103)   &     (0.104)   &     (0.138)   \\
%\addlinespace
\textit{Pizzo}           &               &               &       0.231***&       0.331***&       0.320***&      -0.124   \\
                    &               &               &     (0.081)   &     (0.089)   &     (0.101)   &     (0.101)   \\
%\addlinespace
\textit{PubProc}         &               &               &       0.269** &      -0.156   &      -0.217   &      -0.550** \\
                    &               &               &     (0.137)   &     (0.164)   &     (0.173)   &     (0.254)   \\
%\addlinespace
\textit{Meeting}        &               &               &      -0.046   &      -0.025   &      -0.069   &      -0.218*  \\
                    &               &               &     (0.084)   &     (0.089)   &     (0.082)   &     (0.112)   \\
%\addlinespace
\textit{Boss}         &               &               &               &       0.802***&       0.791***&       1.245***\\
                    &               &               &               &     (0.115)   &     (0.107)   &     (0.181)   \\
%\addlinespace
\textit{Family}          &               &               &               &               &       0.145   &       0.081   \\
                    &               &               &               &               &     (0.110)   &     (0.144)   \\\hline
%\addlinespace 
\textit{Mandamento} FE          &        NO       & NO              &         NO      &     NO          &       NO        &       YES   \\\hline

\midrule
N                   &    9312   &    9312   &    9312   &    9312   &    9312   &    8256   \\
aic                 &    6063.340   &    6013.468   &    5988.480   &    5937.126   &    5937.345   &    5371.396   \\
bic                 &    6077.618   &    6034.885   &    6052.732   &    6008.516   &    6015.874   &    5539.845   \\
\bottomrule
\multicolumn{7}{l}{\footnotesize Standard errors in parentheses}\\
\multicolumn{7}{l}{\footnotesize * p<0.10, ** p<0.05, *** p<0.010}\\
\end{tabular}
}

\caption{Dependent variable=1 if $i$ and $j$ are linked. Estimation is logit, standard errors are bootstrapped with 200 replications. * p$<$0.10, ** p$<$0.05, *** p$<$0.010}
\label{tab:resDyadsFull02newTab5}	
\end{table}

%\begin{table}[H]
%\centering
%\scriptsize
%\input{resDyadsFull02.tex}
%\caption{Dependent variable=1 if $i$ and $j$ are linked. Estimation is logit, standard errors are bootstrapped with 200 replications. * p$<$0.10, ** p$<$0.05, *** p$<$0.010}
%\label{tabDyadFull}	
%\end{table}

In the next section we draw the conclusions from the findings of the social network and econometric analyses on the internal organization of the Sicilian Mafia.

\section{Conclusions\label{secConcl}}
This work analyzed the organizational features of a large Mafia group, drawing on the judiciary evidences of the \textit{Operazione Perseo}, the largest investigation on the Sicilian Mafia in the last 35 years. For the number of individuals and the territory involved, \textit{Operazione Perseo} represents an interesting case study to delve into the organizational features of the Sicilian Mafia.

First of all, the social network analysis highlighted that the \textit{Perseo} network is a small-world network. Specifically, the high clustering is consistent with the recruitment process, which occurs on territorial basis. The econometric analysis, in addition, identifies a decreasing, nonlinear association between the geographical distance and the probability of link formation among agents which, however, loses importance when one controls for the membership of different agents to the same \textit{Mandamento}.

The short diameter in the network can be explained by the need of reaching efficiency, in particular to coordinate Mafia activities across \textit{Mandamenti}, to maintain the rigid demarcation of the territory over which Mafia families rule, to ensure fast circulation of information for the daily conduction of the various illicit trades.

Other aspects of the efficiency of the organization appears from the econometric analysis. In particular, we find that the higher the number of tasks shared by any two agents, the higher the probability that they are connected in the network, as in a multidivision firm \citep{kleinbaum2013discretion}. In addition, the econometric analysis shows that keeping connections across \textit{Mandamenti}, being involved in the collection of extortion money or being a boss implies an increase in the probability for an agent of having links with others. Age, in this respect, does not seem to be particularly significant, pointing out that seniority is of somewhat lower importance in affecting agents' degree.

Efficiency, however, is not the sole criterium to organize a criminal group of Mafia-type. The need for security is also of paramount importance, and is highlighted by several findings. First of all, the small-world properties of the \textit{Perseo} network seem ``adjusted" for its illegal nature. In fact, the clustering coefficient and the diameter are, respectively, lower and higher than in comparable small-world networks of legal nature. In addition, in line with \citet{baccara2008organize}, we retrace in the \textit{Perseo} network a ``cell-dominated hierarchy", in which one key agent is shielded from the rest of the network by sharing a link with another agent (a relative) who plays the role of ``information hub". Familism, of course, implies higher levels of trust among agents, and is another key element of the organizational structure of \textit{Cosa Nostra}.

The hierarchical structure of \textit{Cosa Nostra}, finally, is reflected by the strong disassortativity of the network (a result confirmed by the econometric analysis), a finding in stark contrast to a common feature of social networks, namely, their assortativity. That is, the probability for highly connected individuals is higher to be connected to individuals with low degree. In particular, the econometric analysis shows that if two individuals are bosses (a feature that positively affects their degree), this significantly reduces the probability that they are connected.

Putting these results together, it appears that the Sicilian Mafia is organized following rational principles, as its longevity suggests. Other key aspects of its network structure, namely its connections with non-affiliated members such as professionals, politicians, or businessmen operating in the legal sphere \citep{lavezzi2014organised}, remains an interesting topic for further research.

\newpage

\appendix

\section{Palermo's \textit{Mandamenti}\label{secAMand}}

\begin{figure}[H]
\centering
\includegraphics[scale=0.45, angle=-90]{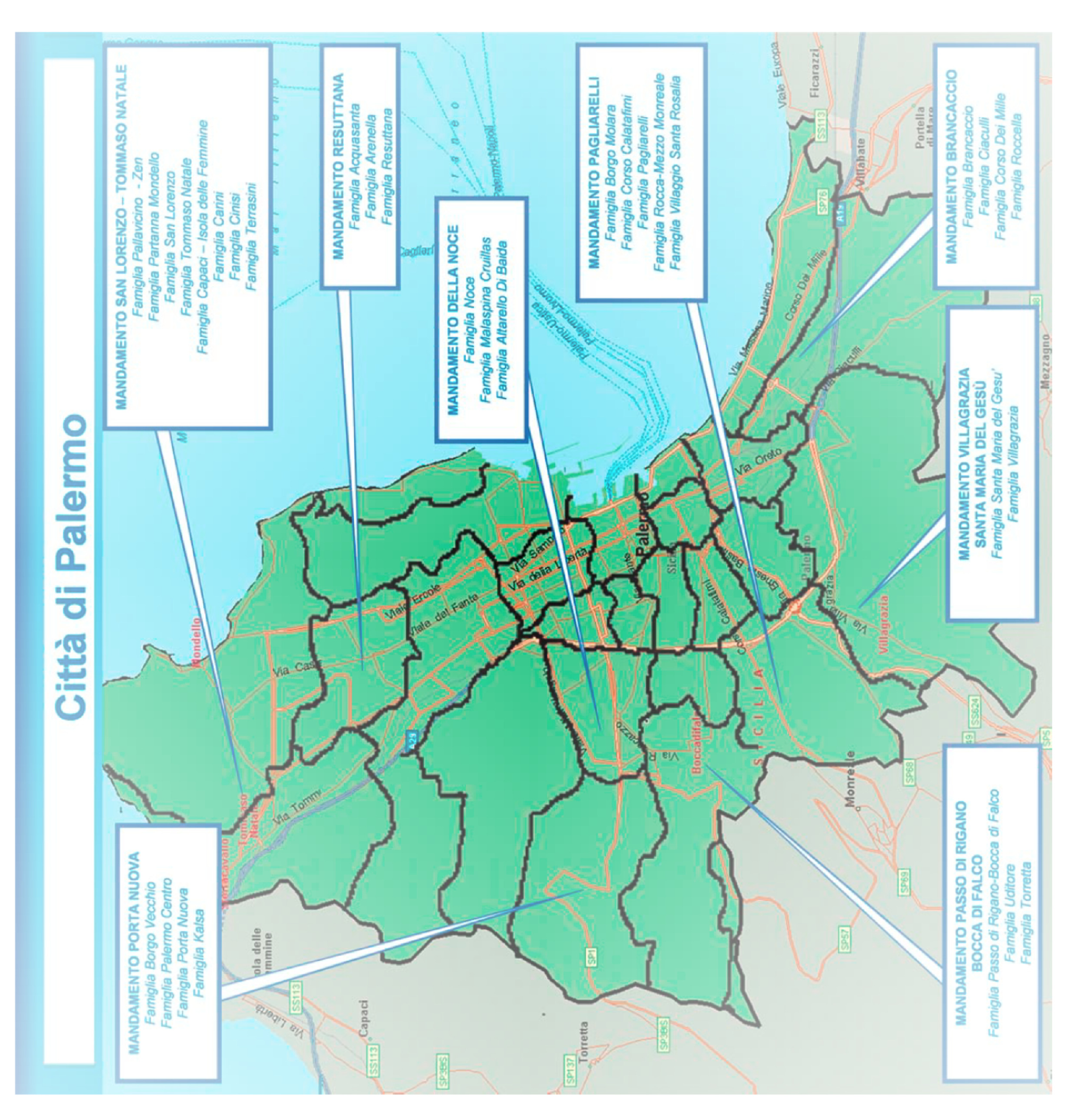}	

\caption{Map of \textit{Mandamenti} in the city of Palermo. Source \cite{dia2016}.}
\label{fig:diaMandamPa}
\end{figure}

\begin{figure}[H]
\centering
\includegraphics[scale=0.45, angle=-90]{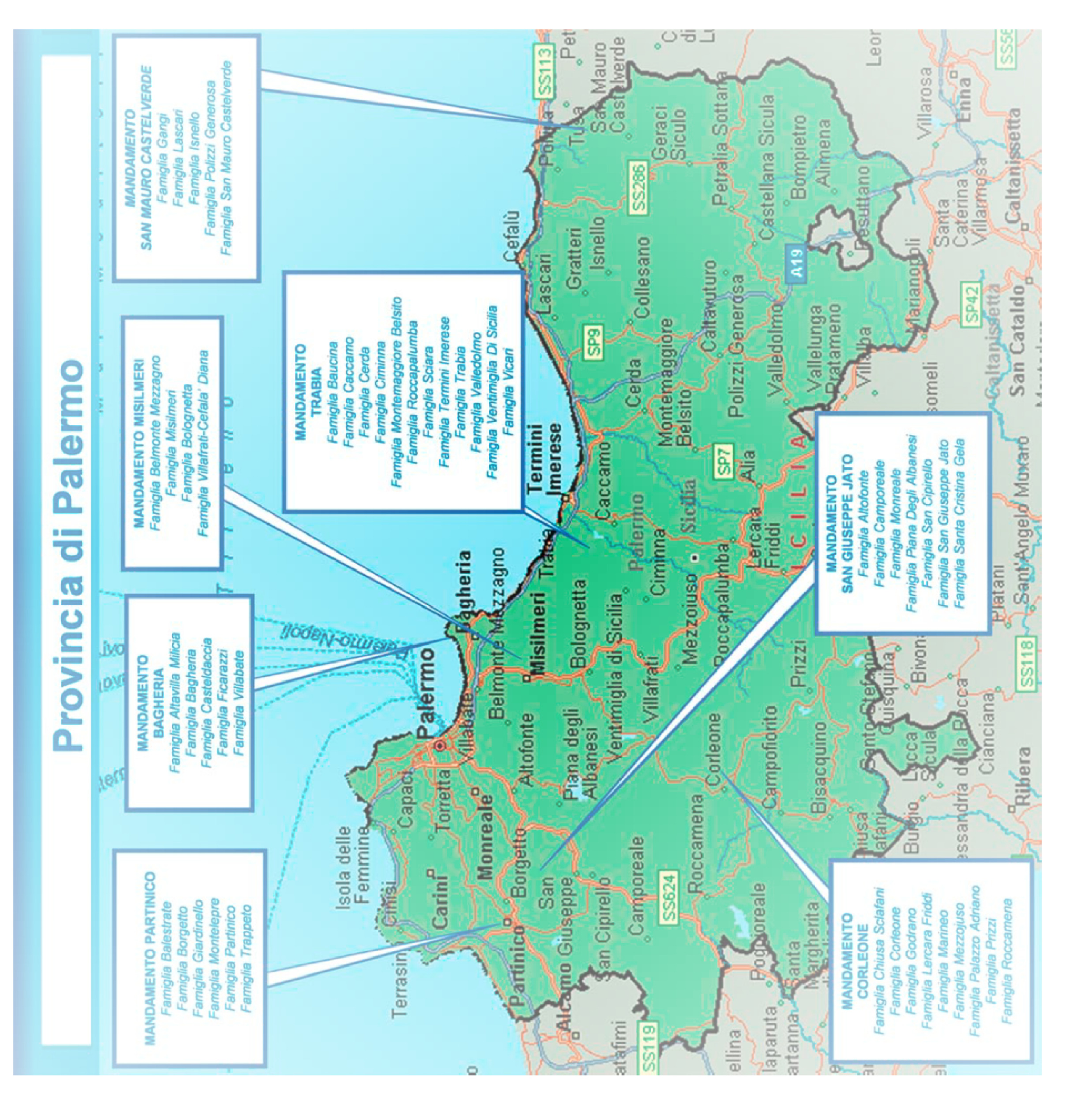}	

\caption{Map of \textit{Mandamenti} in the province of Palermo. Source \cite{dia2016}.}
\label{fig:diaMandamPaPr}
\end{figure}

\section{Results on the Set of 172 agents\label{secA172}}
\begin{figure}[H]
\centering
\includegraphics[scale=0.7,angle=-90]{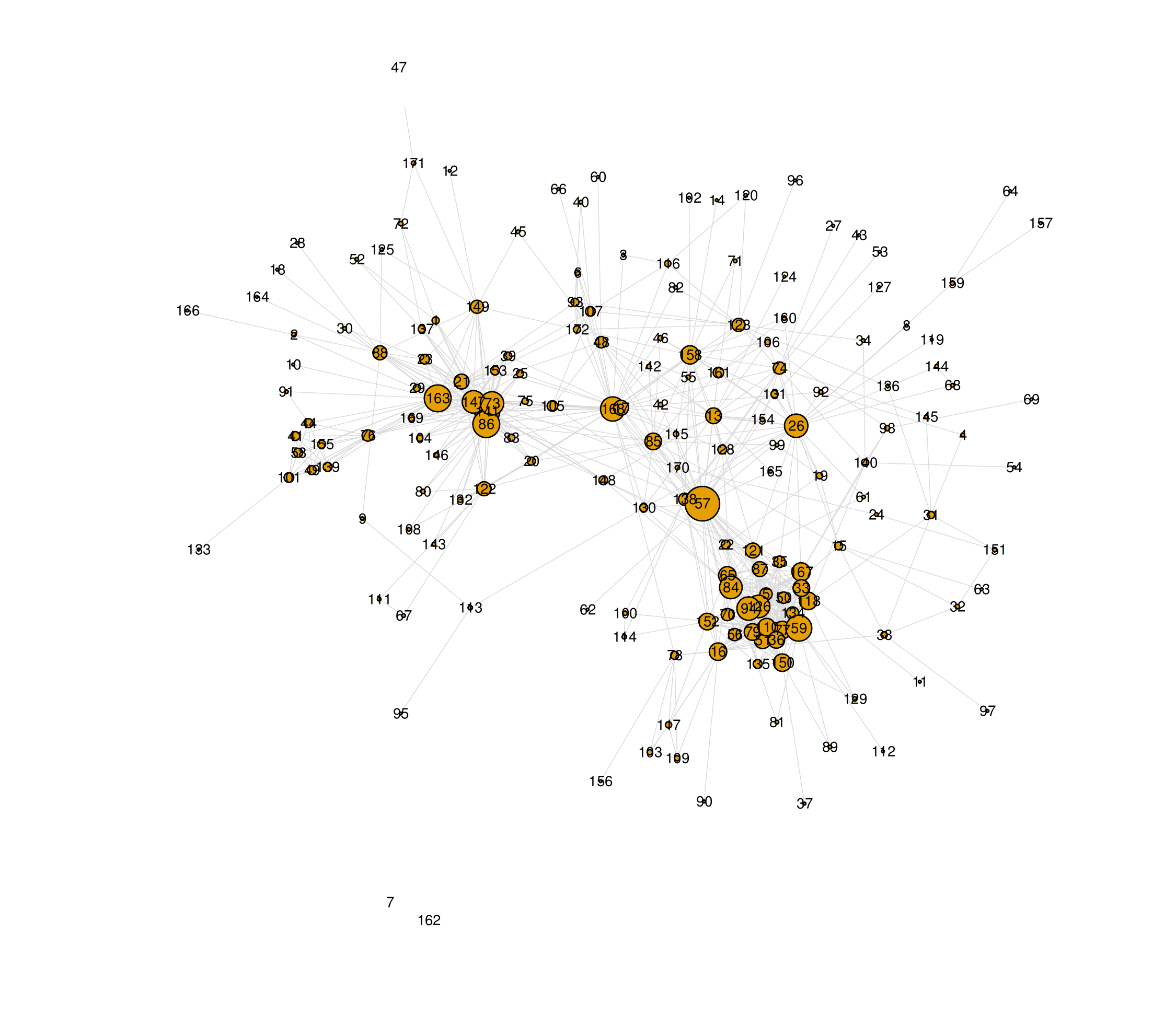}	

\vspace{-1cm}
\caption{Network of the 172 agents identified in the arrest warrant of the \textit{Perseo} operation}
\label{fig:graphNetR}
\end{figure}

Figure \ref{fig:graphNetR} presents the network structure of the whole set of 172 agents identified in the court document examined from \textit{Operazione Perseo}.

Figure \ref{fig:graphNetR} shows that the group of 99 agents of the \textit{Perseo} network represents the core of the network, and the other agents mentioned in the arrest warrant appear as linked to one or few agents of the core group.

\section{Other Results\label{AppOtherRes}}

Table \ref{corrtable} contains the correlations among the tasks of the agents of the \textit{Perseo} network.

\begin{table}[H]\centering 
\begin{tabular}{l  c  c  c  c  c  c  c  c }\hline\hline
\multicolumn{1}{c}{Variables} &Drugs&Guns&dLinkingIN&LinkingOut&Pizzo&PubProc&Meetings&Boss\\ \hline
Drugs&1.000\\
Guns&-0.062&1.000\\
LinkingIn&-0.051&-0.305&1.000\\
LinkingOut&0.080&-0.205&0.538&1.000\\
Pizzo&0.041&-0.169&-0.222&-0.156&1.000\\
PubProc&0.343&-0.084&0.219&0.236&0.259&1.000\\
Meeting&-0.139&-0.120&-0.026&-0.172&-0.230&-0.100&1.000\\
Boss&0.210&-0.123&0.323&0.473&-0.038&0.452&-0.084&1.000\\
\hline \hline 

 \end{tabular}
 \caption{Correlations among Tasks in the \textit{Perseo} Network\label{corrtable}}
\end{table}

\section{The \textit{Perseo} Network is not Scale-Free.\label{app:ScaleFree}}

Given that the \textit{Perseo} network has the small-world properties, we further characterize its properties by checking whether the network has the scale-free property. In a scale-free network the degree distribution takes the form of a power law, that is it can be represented, at least from a certain degree $d$, by a function such as: $P(d)=cd^{-\gamma}$, with $c>0$ and $\gamma>1$ (see, e.g. \citealp[pp. 60-61]{jackson2008social}). The fundamental mechanism to generate a scale-free distribution identified in the literature (see \citealp{barabasi1999emergence}) is based on ``preferential attachment". By this mechanism, the probability that a new node in a growing network connects to an existing node is proportional to the degree of the existing nodes. This generates the fundamental asymmetry in the degree distribution as more connected nodes, at any step of the growth process, have a higher probability of increasing the number of nodes than less-connected nodes.\footnote{\citet{jackson2007meeting} discuss a ``hybrid" model of growing network in which, at any step of the growth process, any new node forms new links in two ways: partially by randomly connecting to existing nodes, and partially by a preferential attachment mechanism.}

\begin{figure}[H]
\begin{minipage}[t]{0.49\linewidth}
	\includegraphics[width=5.5cm,angle=-90]{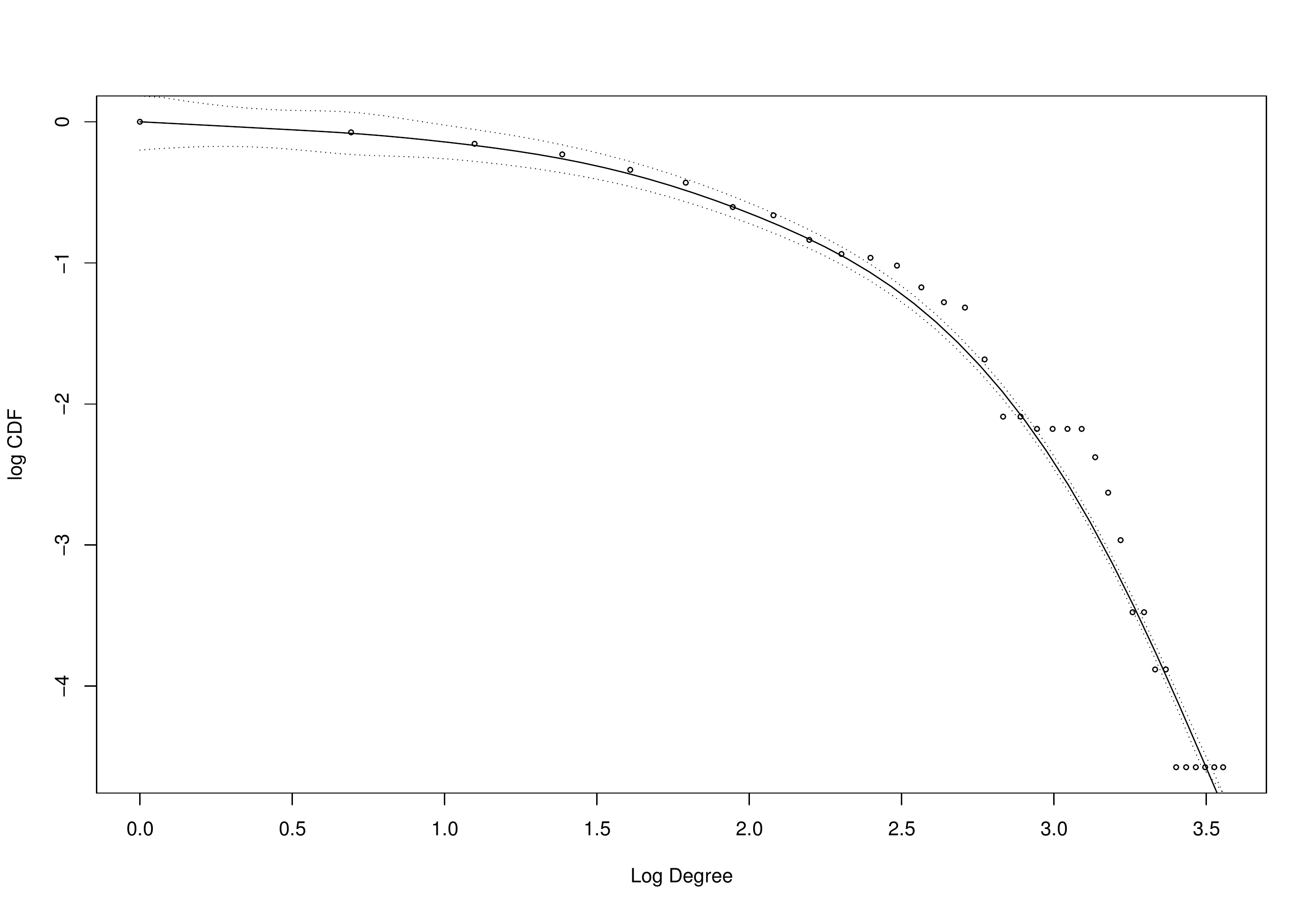}
  \caption{Log(Cumulative distribution of degrees)}
  \label{fig:smRegPL}
    \end{minipage}
\hfill\begin{minipage}[t]{0.49\linewidth}
	\includegraphics[width=5.5cm,angle=-90]{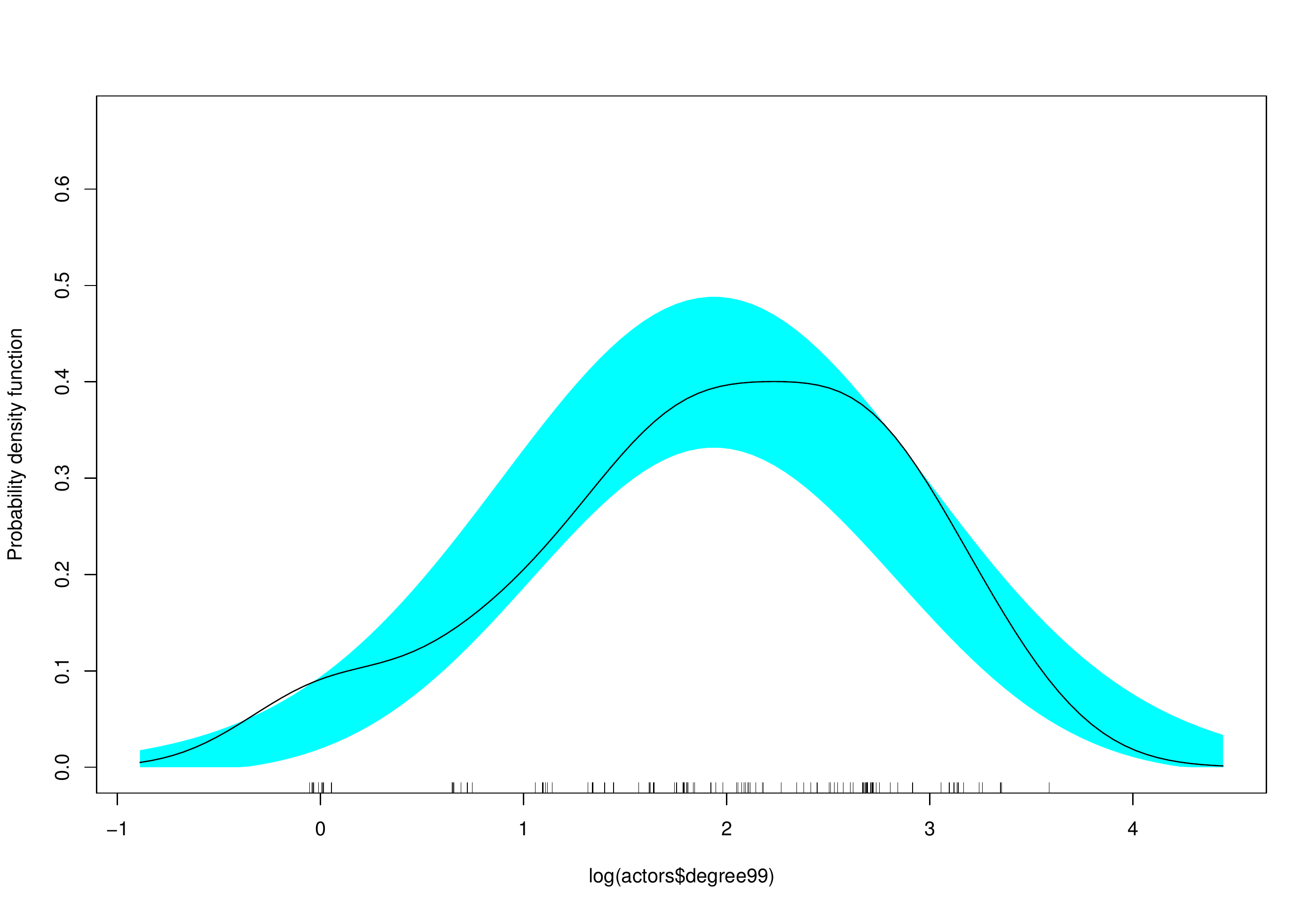}
  \caption{Distribution of Log(Degrees).}
  \label{fig:logNormDegDist}

\end{minipage}
\end{figure}

A way to check for the presence of a scale-free distribution is to plot on a log-log scale the relationship between the cumulative degree distribution and the degree levels. The distribution is scale-free if, at least from  a certain degree level, the relationship is linear with a (negative) coefficient taking on values between 2 and 3.\footnote{See \citet[pp. 132-133]{jackson2008social} for an explanation.}

Figure \ref{fig:smRegPL} shows that the degree distribution is not scale-free. In particular, Figure \ref{fig:smRegPL} highlights that the relationship between the (cumulative) density function and the degree values is far from being linear. On the contrary, the relationship is well-approximated by a concave function\footnote{The function in Figure \ref{fig:smRegPL} is estimated non-parametrically. See \citet[pp.48-52]{bowman1997applied} for details.} with only the last portion, perhaps, looking linear. Indeed, if we estimate the presence of a power-law relationship following the method proposed by \citet{clauset2009power}, we find that there exists a negative linear relationship starting from a degree value of 23 (3.13 in logs), with a coefficient of -8.20.\footnote{This coefficient is highly significant. This estimation is performed using the function \texttt{fit\_power\_law} in the \textit{R} package \texttt{igraph}.} However, as pointed out by \citet{broido2019scale}, such a coefficient comes from processes most likely classifiable as ``weak/weakest/super-weak" scale-free.\footnote{Intuitively, a negative relationship with such a high coefficient implies that the probability of having agents with a relatively high degree decays much faster than what implied by a coefficient comprised between 2 and 3.}  

Figure \ref{fig:logNormDegDist}, instead, shows that the concave relationship estimated in Figure \ref{fig:smRegPL} corresponds to a log-normal degree distribution.\footnote{\citet[p. 243]{mitzenmacher2004brief} shows that, indeed, a lognormal distribution is associated to a concave relationship between the CDF and the values of a random variable, in a log-log form.} In particular, Figure \ref{fig:logNormDegDist} shows that we do not reject the null hypothesis that the degree distribution is log-normal.\footnote{In particular, the estimation lies within the variability bands, constructed as suggested by \citet[pp. 75-76]{bowman1997applied}, by adding two standard errors above and below the point estimates of the density function, which is estimated non-parametrically, with normal optimal smoothing.} 

\section{Robustness Tests\label{app:RobustTest}}
% A) 416 bis;  B) 416 bis; C) Droga; D) Droga; E) - S) reati vari, tra cui minacce, estorsioni, favoreggiamento
The first section of the arrest warrant (22 pages) contains the list of the individuals apprehended and a summary of the accusations (\textit{capi d'imputazione}) against them. The accusations refer to crimes such as: i) being a member of the Mafia; ii) drug smuggling; iii) threats; iv) extortion; v) providing support to other criminals.

In listing the accusations, the Prosecutor also provides information that can be utilized to infer the existence of a link between an individual and others implied in the investigation. As pointed out in Footnote \ref{footIntroDoc}, such pieces of information have a form as: ``[agent 110] ... substituted [agent 77] in the direction of the Mafia family ... and kept contacts with members of the XYZ \textit{Mandamento}." From such a piece of information we assumed that there exists a link between agents 110 and 77. The information on the connections between an agent and the members of a certain \textit{Mandamento} identified in this way, that we denote as a ``generic links", posed us in front of a dilemma.

On the one hand, the information on the existence of ``generic links" is of high quality as, in the very words of the Prosecutor, these links existed and, therefore, it is not necessary to resort to other pieces of evidence such as wiretapped conversations and the like. In addition, it represents key information on connections across \textit{Mandamenti}, an important but rather unexplored issue in the network analysis of \textit{Cosa Nostra}. On the other hand, the document does not report the exact identification of \textit{which} members of the \textit{Mandamento} shared links with the agent. There are at least two straightforward choices about ``generic links" that can be made in \textit{mapping} the network \citep{bichler2019understanding} from the document: ignoring them altogether, as there is no exact identification of the connected agent(s), but the indication of their \textit{Mandamento} only, therefore giving up valuable information, or utilizing them, as the document still contains some hints on how to identify the connected agents, i.e. the \textit{Mandamento} affiliation (see Footnote \ref{foot:MandamDistib} for the distribution of agents across \textit{Mandamenti}). In the latter case, the question is how many links among the agent and those in the court document belonging to that \textit{Mandamento} exist.

Our final choice was to assume that a ``generic link" implies that the agent involved has connections with \textit{all} the agents from a certain \textit{Mandamento} that are mentioned in the court document. This choice has the advantage of allowing a straightforward definition of the links, but leaves open the question of whether we are overestimating the number of existing links, as we cannot be sure that an agent was actually connected to \textit{all} the agents of a \textit{Mandamento} mentioned in the arrest warrant.

\begin{figure}[H]
\begin{minipage}[t]{0.49\linewidth}
	\includegraphics[width=5.5cm,angle=-90]{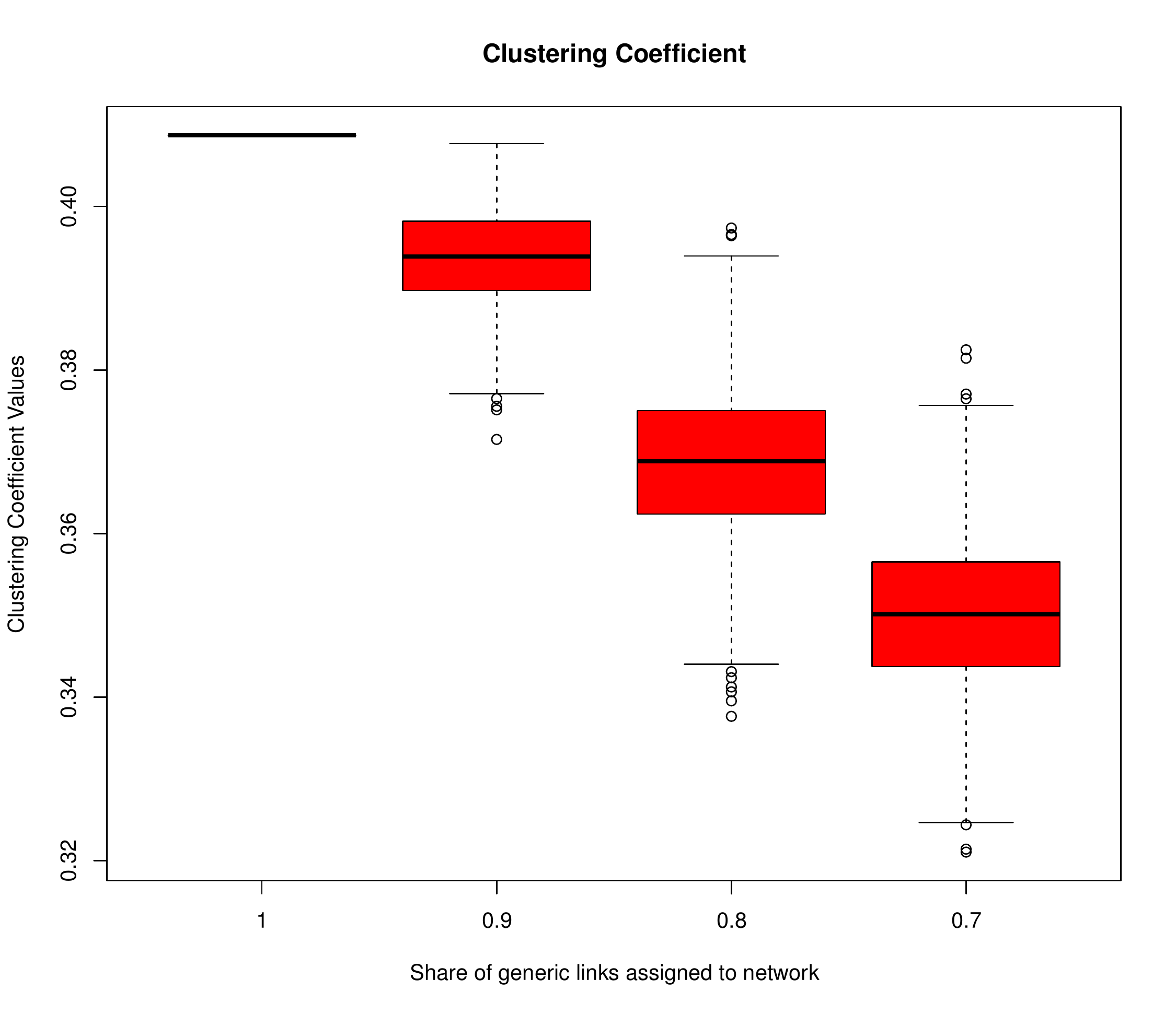}
  \caption{Box plot of the Clustering Coefficient values obtained from simulated networks}
  \label{fig:boxPClustCoeff}
    \end{minipage}
\hfill\begin{minipage}[t]{0.49\linewidth}
	\includegraphics[width=5.5cm,angle=-90]{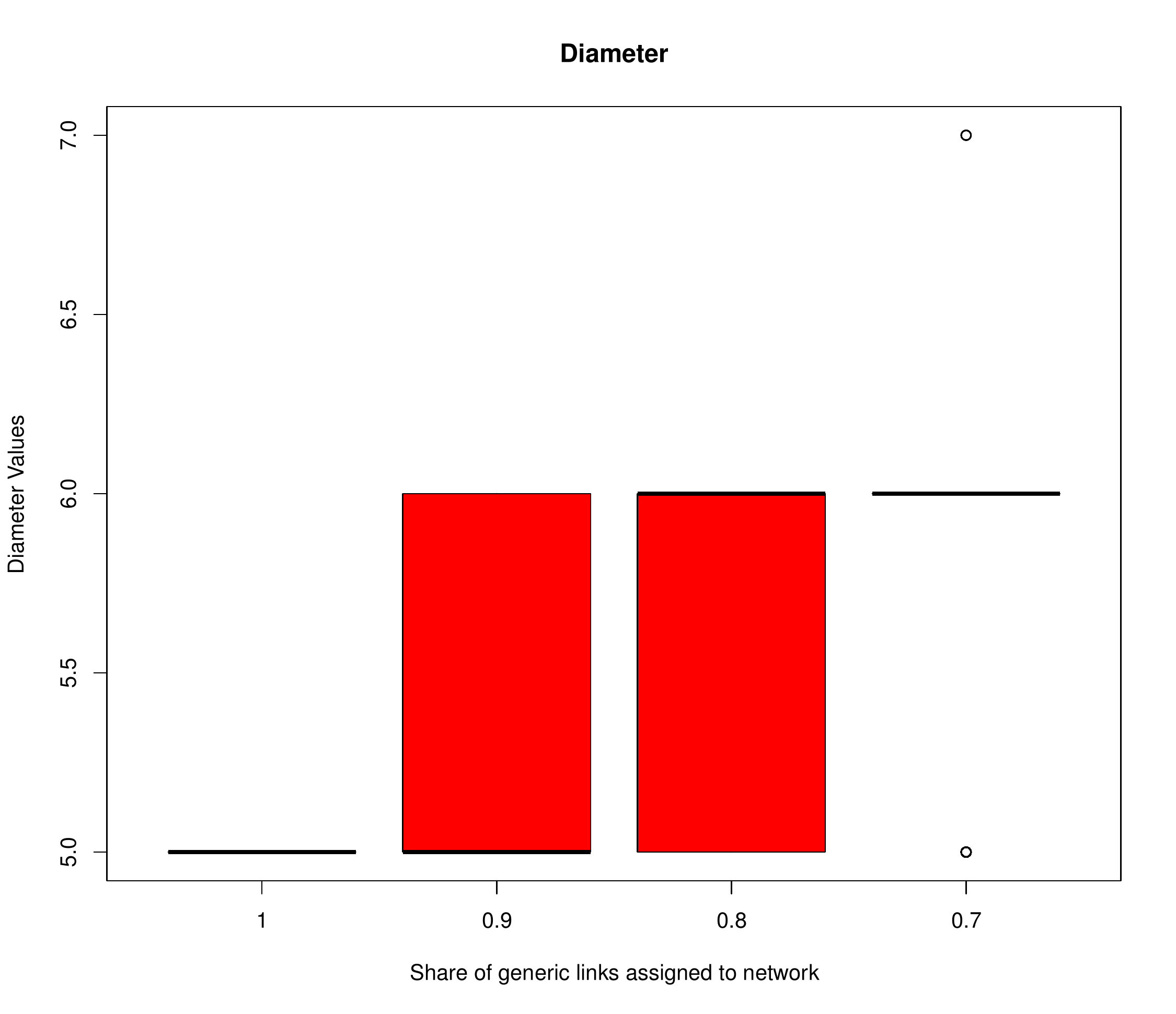}
  \caption{Box plot of the Diameter values obtained from simulated networks}
  \label{fig:boxPDiamter}

\end{minipage}
\end{figure}

\begin{figure}[H]
\begin{minipage}[t]{0.49\linewidth}
	\includegraphics[width=5.5cm,angle=-90]{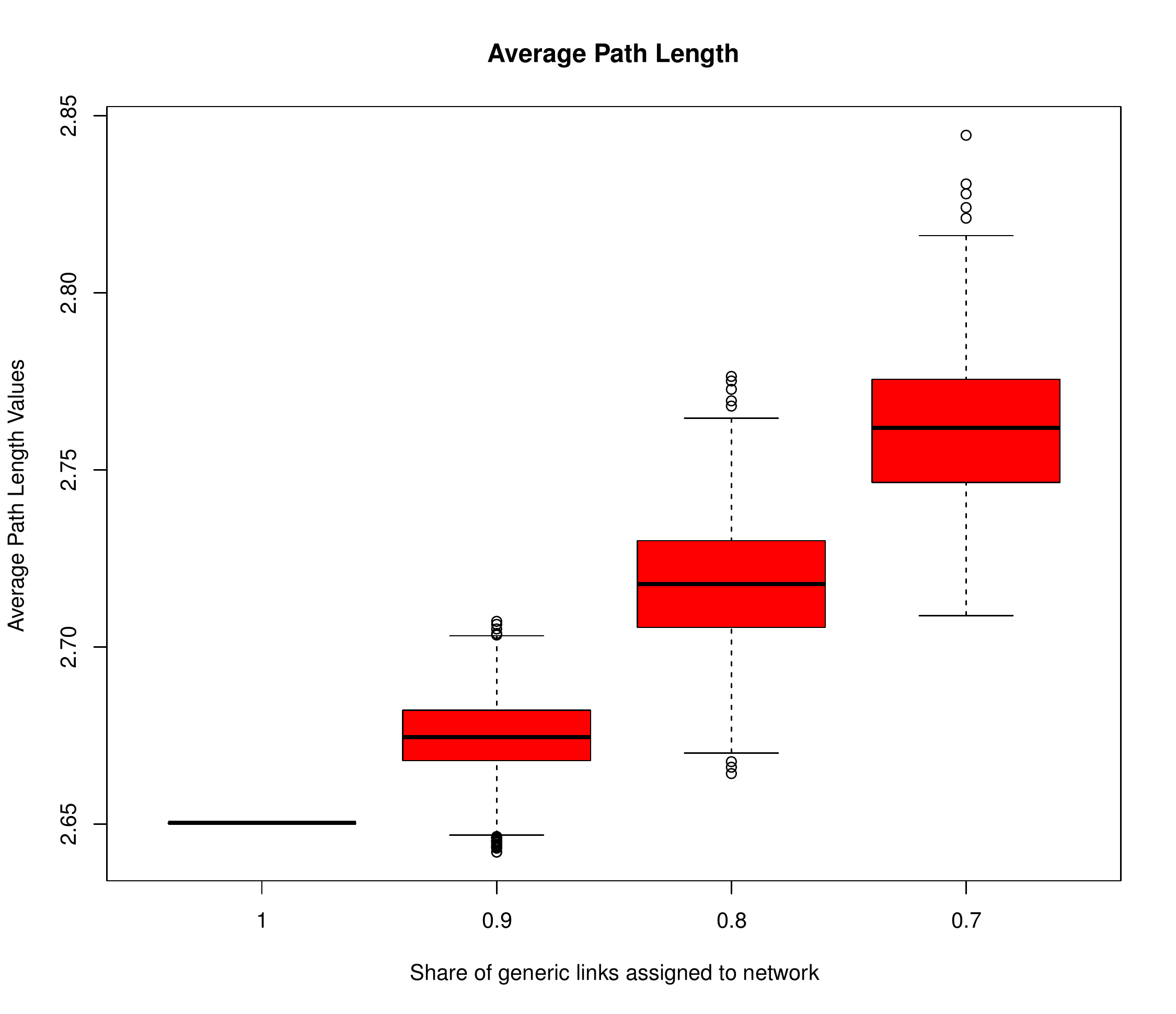}
  \caption{Box plot of the Average Path Length values obtained from simulated networks}
  \label{fig:boxPAPL}
    \end{minipage}
\hfill\begin{minipage}[t]{0.49\linewidth}
	\includegraphics[width=5.5cm,angle=-90]{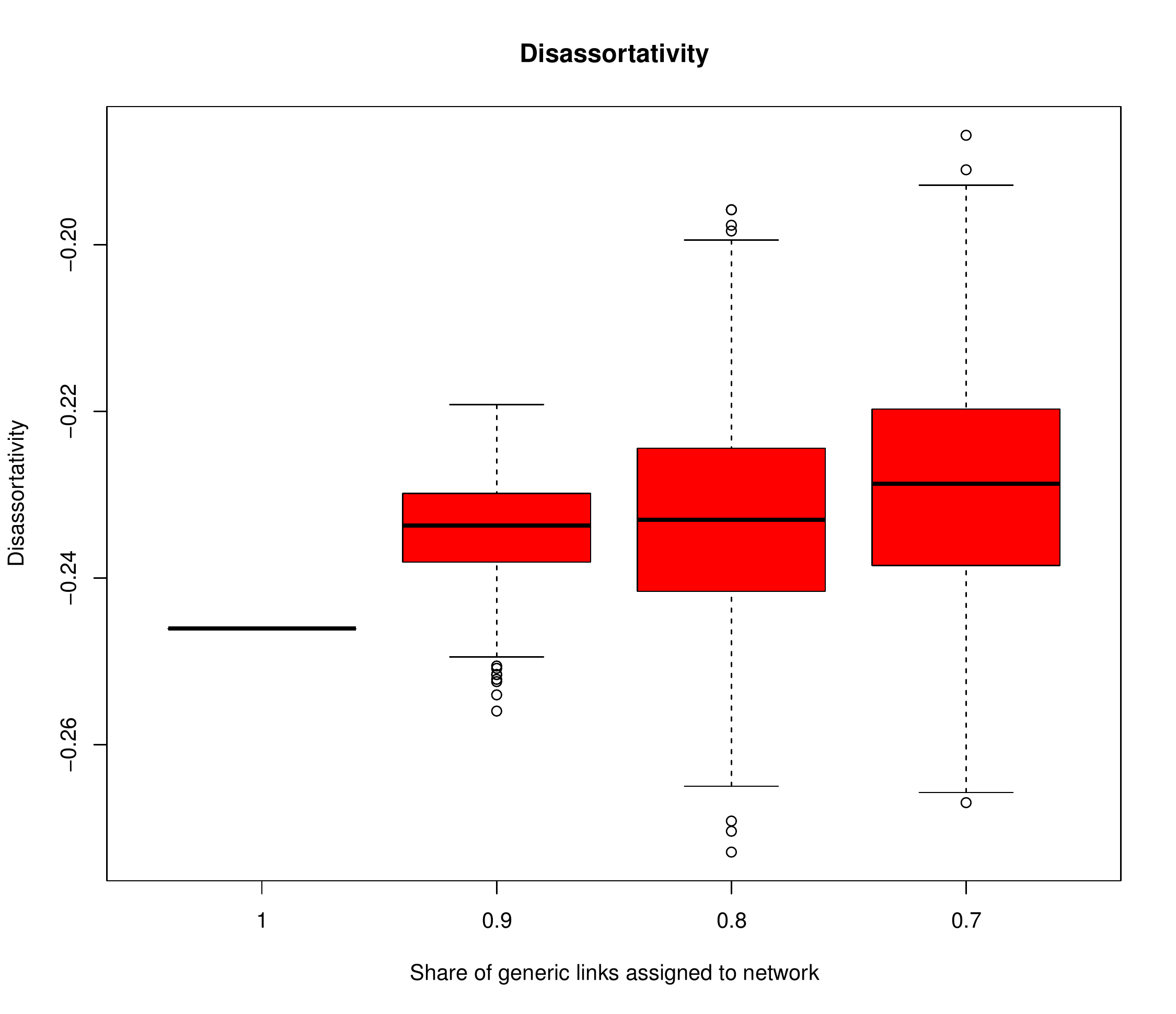}
  \caption{Box plot of the Assortativity values obtained from simulated networks}
  \label{fig:boxPDis}

\end{minipage}
\end{figure}

To take into account this possible shortcoming we adopted the following strategy. First of all, we checked which ``generic links" are actually confirmed by other pieces of evidence retrieved in the second part of the document. The remaining ``generic links" were the subject of simulations in which the simulated networks had different shares of ``generic links" that were randomly turned into ``confirmed links". For example, in a simulation in which we assumed that 90$\%$ of ``generic links" actually exist, for each agent in the dataset having ``generic links" with members of a \textit{Mandamento}, we randomly created a link with a share of 90$\%$ of the members of that \textit{Mandamento} in the dataset. For each share (90$\%$, 80$\%$, etc.) we simulated 1000 networks and computed the main topological characteristics considered in this paper, namely: i) the clustering coefficient; ii) the diameter; iii) the average path length; iv) the assortativity level (see Section \ref{secComplexNet} for details on the definitions of these measures). Figures \ref{fig:boxPClustCoeff}, \ref{fig:boxPDiamter}, \ref{fig:boxPAPL}, and \ref{fig:boxPDis} summarize the results of the simulations.

The random assignment of a decreasing share of links has some consequences by construction. First of all, the number of links decreases. In particular, with respect to the 470 links existing in the \textit{Perseo} network, the (average) number of links with the random assignment of, respectively 90$\%$, 80$\%$, and  70$\%$ of ``generic links" is 449, 416, 388. In addition, with a sparser network, the distance across nodes is likely to increases. Figures \ref{fig:boxPDiamter} and \ref{fig:boxPAPL} confirm that this is the case. The diameter increases (in median value) from 5 to 6, while the average path length increases (in median value) from 2.65 in the \textit{Perseo network} to, respectively, 2.67, 2.72, and 2.76. Finally, the clustering coefficient decreases. This also should be expected, as the random assignment of ``generic links" is referred to individuals with the same \textit{Mandamento} affiliation who are likely to be highly interconnected. A decreasing share of links with interconnected agents implies a decrease in the clustering coefficient of an agent. However, with respect to these topological measures, we see that the distribution of values across different simulations are very often largely overlapping.

Interestingly, a key parameter shows a remarkably high robustness across simulations: the disassortativity level. In fact, the value computed for the \textit{Perseo} network ($-0.25$) is always very similar to the median values (approximately $-0.23$) computed in the different simulations and strictly included in the distribution of values of the different simulations, which always show a large overlapping of the interquartile ranges.

In addition, for the simulated networks of this robustness test, the results of Section \ref{secComplexNet} based on comparisons with random networks and Small World networks  I and II are confirmed. In particular, the simulated criminal networks are small-world, i.e. they have a higher clustering coefficient a small diameter, when observed with respect to comparable random networks. In addition, when observed with respect to comparable Small World networks with similar clustering coefficient, they have a higher diameter, while with respect to comparable Small World networks with similar diameter, they have a lower clustering coefficient (results are available upon request).

%{\normalsize 
%\bibliographystyle{chicago}
%\bibliography{sample}
%}

\bibliographystyle{chicago}
\bibliography{snArxiv01}

\end{document}